\documentclass[12pt,draftcls,onecolumn]{IEEEtran}

\usepackage[text={17cm,24cm}]{geometry}

\usepackage{booktabs}

\usepackage{amsmath,amssymb,subfigure,graphicx,xcolor,color,transparent}
\newtheorem{problem}{Problem}[section]

\newtheorem{remark}{Remark}[section]
\newtheorem{assumption}{Assumption}
\newtheorem{lemma}{Lemma}[section]
\newtheorem{theorem}{Theorem}[section]
\newtheorem{corollary}{Corollary}[section]

\newenvironment{proof}[1][Proof]{\textbf{#1.} }{\ \rule{0.5em}{0.5em}}

\begin{document}
\title{Lyapunov stochastic stability and control of robust dynamic coalitional games with transferable utilities
\thanks{Preliminary conference versions of this work were presented in Allerton 2010 \cite{BR10A} and CDC 2010 \cite{BR10}. The authors would like 
to thank Ehud Leher and Eilon Solan for their support in exploring connections with approachability and attainability.}}
\author{Dario Bauso,\thanks{D. Bauso is with Dipartimento di Ingegneria Chimica, Gestionale, Informatica e Meccanica, 
Universit\`a di Palermo, Italy, email: \textsl{dario.bauso@unipa.it}} \and 
P. Viswanadha Reddy,\thanks{P.V. Reddy is with GERAD and HEC Montreal, email: \textsl{viswanadha.puduru@gmail.com}} \and
Tamer Ba\c{s}ar,\thanks{T. Ba\c{s}ar is with Coordinated Science Laboratory, University of Illinois at
Urbana-Champaign, Urbana, IL, USA, e-mail: \textsl{basar1@illinois.edu}}}

%
%
%
%
\maketitle
\begin{abstract}
This paper considers a dynamic game with transferable utilities (TU), where the characteristic function is a
continuous-time bounded mean ergodic process. 
A central planner interacts continuously over time with the players by choosing 
the instantaneous allocations subject to budget constraints.
Before the game starts, the central planner knows the nature of the process (bounded mean ergodic), 
the bounded set from which  the coalitions' values are sampled, and the long run average coalitions' values.
On the other hand, he has no knowledge of the underlying probability function generating the coalitions' values. 
Our goal is to find allocation rules that use a measure of the extra reward that a coalition has received up to the current time by re-distributing the budget among the players. 
The objective is two-fold: i) guaranteeing convergence of the average allocations to the core (or a specific point in the core)
of the average game, ii) driving the coalitions' excesses to an {\it a priori} given cone.
The resulting allocation rules are \textit{robust} as they guarantee the aforementioned convergence properties 
despite the uncertain and time-varying nature of the coaltions' values. 
We highlight three main contributions. 
First, we design an allocation rule based on full observation of the extra reward so that 
the average allocation approaches a specific point in the core of the average game, while
the coalitions' excesses converge to an {\it a priori} given direction. 
Second, we design a new allocation rule based on partial observation on the extra reward
so that  the average allocation converges to the core of the average game, while
the coalitions' excesses converge to an {\it a priori} given cone.  
And third, we establish connections to approachability theory \cite{B56,L02} and attainability theory \cite{BLS12,LSB11}.

\end{abstract}


\section{Introduction}\label{Introduction}
Coalitional  games with transferable utilities (TU), introduced first by Von Neuman and Morgenstern \cite{VM}, have recently sparked much interest in the control and communication engineering communities \cite{SHDHB09}. In essence, coalitional TU games are comprised of a set of players who can form coalitions and a characteristic function associating a real number with every coalition. This real number represents the value of the coalition and can be thought of as a monetary value that can be distributed among the members of the coalition according to some appropriate fairness allocation rule. The value of a coalition
also reflects the monetary benefit demanded by a coalition to be a part of the grand coalition.  

This paper considers a \textit{dynamic TU game}, where the characteristic function is a
bounded mean ergodic process. Bounded means that the characteristic function
takes values in a convex set according to an unknown probability
distribution. Mean ergodic means that the expected value of the coalitions values at each time coincides with the
long term average.
With the dynamic game we associate a {\it dynamic average game} obtained by averaging over time the coalitions' values,
and assume that the core of the average game is nonempty on the long run. 
Given the above dynamic TU game, a central planner interacts continuously over time with the players by choosing 
the instantaneous allocations subject to budget constraints.
Before the game starts, the central planner knows the nature of the process (bounded mean ergodic), the bounded set  and the long run average coalitions' values.
On the other hand, he has no knowledge of the underlying probability function generating the instantaneous coalitions' values. 
Our goal is to find allocation rules that use a measure of the extra reward that a coalition has received up to the current time by re-distributing the budget among the players. 
The objective is two-fold: i) guaranteeing convergence of the average allocations to the core (or a specific point in the core)
of the average game, ii) driving the coalitions' excesses to an {\it a priori} given cone.
The resulting allocation rules are \textit{robust} as they guarantee the aforementioned convergence properties 
despite the uncertain and time-varying nature of the coaltions' values. 

In the context of coalitional TU games, \textit{robustness} and \textit{dynamics} naturally arise in all the situations where the coalitions values are uncertain and time-varying, see e.g., \cite{BT09}. 
Robustness has to do with modeling coalitions' values as unknown entities and this is in spirit with some literature on stochastic coalitional games \cite{SB99,TBT03}. However, we deviate from the latter works since the probability function
generating the random coalitions values is unknown, and this is more in line with the concept of 
Unknown But Bounded (UBB) variables formalized in \cite{B71}. It is worth to mention that this formulation shares some common elements with the recent literature on interval valued games \cite{AMT}, where the authors use intervals to describe coalitions values quite similar to what is done in this paper. The interval nature of coalitions' values arises generally due to the optimistic and pessimistic expectations of the coalitions \cite{LBIA08} when cooperation is achieved from a strategic form game. We also note some differences in that we focus here more on the time-varying nature of the coalitions' values. 
In doing so, we also link the approach to the set invariance theory \cite{B99} and stochastic stability theory \cite{LF96} which provides us some \textit{nice} tools for stability analysis (see, e.g., the use of a Lyapunov function in the proof of Theorem \ref{thm1}). 

Bringing dynamical aspects into the framework of coalitional TU games is an element in common with other papers \cite{FP00,H75,KPP05}. The main difference with those works is that the values of coalitions are realized exogenously and no relation exists between consecutive samples. 

Convergence conditions together with the idea that allocation rules use a measure of the extra reward that a coalition has received up to the current time by re-distributing the budget among the players are a main issue in a number of other papers \cite{AS02,C98,G76,L02,SS96} as well. However, this paper departs from the aforementioned ones mainly in that dynamics in those works is captured by a bargaining mechanism with fixed coalitions' values while we let the values be time-varying and uncertain. This last element adds some robustness to our allocation rule which has not been dealt with before.   
%


The main contribution of this paper is captured by the following three results. 
First, we design an allocation rule based on full observation of the extra reward so that 
the average allocation approaches a specific point in the core of the average game, while
the coalitions' excesses converge to an  {\it a priori} given direction. 
Second, we design a new allocation rule based on partial observation on the extra reward
so that  the average allocation converges to the core of the average game, while
the coalitions' excesses converge to an {\it a priori} given cone.  
Convergence of both allocation rules is proved via Lyapunov stochastic stability theory. And third, we establish connections of the Lyapunov stochastic stability theory to the approachability theory \cite{B56,L02} and attainability theory \cite{BLS12,LSB11}.

A few other contributions of the paper are the definition of average game, whose role becomes fundamental
when the coalitions' values variations are known with delay by the planner; the reformulation of the problem as
a network flow control problem, where the allocation rule turns into a robust control policy is a novel aspect, with the importance of such a reformulation lying in the fact that we can prove the convergence of the allocations 
using the strong tools of the Lyapunov stochastic stability theory; 
and finally, the idea of turning a coalitional TU game set up into a control theoretic 
problem is a novel one, which represents, by far, the main characteristics of this work.

%
%
%
The paper is organized as follows. In Section \ref{sec:pf}, we formulate the problem. In Section \ref{sec:FT},
we present the basic idea of our solution approach. In Section \ref{sec:mainresults} 
we state the three main results of this work and postpone the derivation of such results to Section \ref{sec:resultsderiv}. 
In Section \ref{sec:sim}, we provide some numerical illustrations. 
Finally, in Section \ref{sec:conclusions}, we draw some concluding remarks. 

\noindent
{\bf Notation}. 
We view vectors as columns. For a vector $x$, we use $x_i$ or $[x]_i$ to denote
its $i$th coordinate component. 
For two vectors $x$ and $y$, we use $x<y$ ($x\le y$) to denote
$x_i<y_i$ ($x_i\le y_i$) for all coordinate indices $i$. 
We let $x^T$ denote the transpose of a vector $x$, and $\|x\|_n$ 
denote its $n$-norm.  
For a matrix $A$, we use $a_{ij}$ or $[A]_{ij}$
to denote its $ij$th entry. We use $|a_{ij}|$ to denote the absolute value
of scalar $a_{ij}$.
Given two sets $U$ and $S$, we write $U\subset S$ to denote that $U$
is a proper subset of $S$.
We use $|S|$ for the cardinality of a given finite set $S$.
Let  $\Phi$ be a closed and convex set in $\mathbb R^m$, we use $P(y)$ to denote the projection of any point 
$y \in \mathbb R^m$ onto $\Phi$ (closest point to $y$ in $\Phi$). 
We also denote by $\partial \Phi$ the boundary of $\Phi$ and 
$n_{y}$ the outward normal for any $y\in \partial \Phi$.
We use $dist(y,\Phi)$ to denote the euclidean distance between point $y$ and set $\Phi$.
Given a set $N$ of players and a function $\eta: S\mapsto
\mathbb{R}$ defined for each nonempty coalition $S\subseteq N$, 
we write $<N,\eta>$ to denote the transferable utility (TU) game with 
the players' set $N$ and the characteristic function $\eta$.  
We let $\eta_S$ be  the value $\eta(S)$ of the characteristic 
function $\eta$ associated with a nonempty coalition $S\subseteq N$.
Given a TU game $<N,\eta>$, we use $C(\eta)$ to denote the core of the game, 
$C(\eta)=\left\{x \in\mathbb{R}^{|N|} \,\Big|\, \sum_{i\in N}x_i=\eta_N,\  
\sum_{i\in S} x_i\ge\eta_S \hbox{ for all nonempty } S\subset N\right\}.$
Also, $\mathbb{R}_{+}$ denotes the set of nonnegative real numbers.
Given a random vector $\xi$ the notation $\mathbb E[\xi]$ denotes its expected value. 
Given a random process $\{v(t)\}$ we denote by 
$\tilde v(t) = \int_0^t  v(\tau) d\tau$, its integral and 
$\bar v(t)=\frac{\tilde  v(t)}{t}$ its average up to time~$t$.

\section{Model and problem formulation}
\label{sec:pf}

In this section, we formulate the problem in its generic form and elaborate on the role of information. 
Let $N=\{1,\ldots,n\}$ be a set of players and $S\subseteq N$ the set of all
(nonempty) \emph{coalitions}  arising among these players.
Denote by $m=2^n-1$ the {\it number of possible coalitions}.
We assume that time is continuous and use $t\in \mathbb R_+$ to index the
time slots.

We consider a \textit{dynamic TU game}, denoted $<N,\{v(t)\}>$, where $\{v(t)\}$ is a
continuous flow of characteristic functions. The flow $\{v(t)\}$ describes a bounded mean
ergodic process. By bounded we mean that given a bounded convex set $\mathcal V \in \mathbb R^m$ and  
a probability function $\mathbb P \in \Delta(\mathcal{V})$, where $\Delta(\mathcal V)$
is the set of probability functions on $\mathcal V$, 
then  for all $t \in \mathbb R_+$ each random variable $v(t)$ takes values in $\mathcal V \in \mathbb R^m$
according to probability $\mathbb P$ 
as expressed in (\ref{bound}); by mean ergodic we mean that 
its expected  value coincides with the long term average
as in (\ref{ergo}): 
\begin{eqnarray}\label{bound} v(t)\in \mathcal V \subset \mathbb R^m,\quad \mbox{for all } t \in \mathbb R_+\\\label{ergo} \mathbb E[v(t)] = lim_{\tau\rightarrow \infty} \bar v(\tau),\quad  \mbox{for all } t \in \mathbb R_+.\end{eqnarray}

Thus, in the dynamic TU game $<N,\{v(t)\}>$, the players 
are involved in a sequence of instantaneous TU games whereby, at 
each time $t$, the {\it instantaneous TU game} is
$<N,v(t)>$ with $v(t)\in \mathcal V$ 
for all $t\ge0$. Further, we let $v_S(t)$ denote {\it the value assigned to
a nonempty coalition} $S\subseteq N$ in the instantaneous game $<N,v(t)>$.



With the dynamic game we associate a {\it dynamic average game} $<N,\{\bar v(t)\}>$ and
an {\it instantaneous average game at time $t\ge 0$}, $<N,\bar v(t)>$. 

%
The motivation of formalizing the above dynamic TU games is in that such games represent a stylized model of all those scenarios where the coalitions' values vary with time. 


%

We assume that the core of the average game is nonempty on the long run. We will see that without this assumption 
the problem under study has no solution. 
Thus, denote by $v_{nom}$ the (long run) average coalitions' values, namely, $v_{nom}:=\lim_{t \rightarrow \infty} \bar v(t)$ 
and let $C(v_{nom})$ be the core of the average game. 
\begin{assumption}\label{asm:ave} \textbf{(balancedness)}
The core of the average game is nonempty in the limit:
$C(v_{nom}) \not = \emptyset$.
\end{assumption}
We can view the above assumption as introducing some steady-state (average) conditions on a
game scenario subject to instantaneous fluctuations. 
However, note that we do not make assumptions regarding the balancedness of the instantaneous
games which is the case with \cite{BT09}. 
Thus, the core of the instantaneous game can be empty at some time $t$. 

Given the above dynamic TU game, a central planner interacts continuously over time with the players by choosing 
the instantaneous allocations denoted by $a(t) \in \mathbb R^n$. We assume that the allocations 
are subject to the following budget constraints.

\begin{assumption}\label{asm:bounded} \textbf{(bounded allocation)}
The instantaneous allocation is bounded within a hyperbox in~$\mathbb R^n$
$$a(t) \in \mathcal A := \{a\in \mathbb R^n:\, a_{min} \leq a \leq a_{max}\},$$
with {\it a priori} given lower and upper bounds $a_{min}$, $a_{max} \in \mathbb R^n$.
\end{assumption}

As regards the information available  {\it a priori} (before the game starts) to the central planner, we assume that
he knows the nature of the process $\{v(t))\}$ (bounded mean ergodic), the bounded set $\mathcal V$ and the long run average coalitions' values
$v_{nom}$. The latter is the same as saying that he knows the expected coalitions' values for all $t\in \mathbb R_+$. 
On the other hand, he has no knowledge of the underlying probability function $\mathbb P$. 
%
\begin{assumption} \label{ass:info} \textbf{(on available information)}
The planner knows $v_{nom}$.
\end{assumption}
\noindent

Beside this, during the game the central planner also observes the extra reward of the coalitions up to $t$ and for all $t\in \mathbb R_+$.
Given this, and in line with a number of other papers \cite{AS02,C98,G76,L02,SS96}, our goal is to find allocation rules that use a measure of the extra reward that a coalition has received up to the current time by re-distributing the budget among the players. To do this, a first step  is to define excesses for the coalitions.
For any coalition $S \subseteq N$, we
define \emph{excess (extra reward) at time $t \geq 0$} as the \emph{excess at time $t = 0$} plus the
difference between the total integral reward, given to it, and the integral value of the coalition itself, i.e.,
$${\epsilon}_S(t)= \sum_{i\in S} \tilde a_i(t) - \tilde v_S(t) + {\epsilon}_S(0).$$ Furthermore, assuming without loss of generality ${\epsilon}_S(0)=0$, we say that \emph{$S$ is in excess at time $t \geq 0$} if the excess is nonnegative, i.e.,
$\sum_{i\in S} \tilde a_i(t) \geq \tilde v_S(t)$. 
%
Let $\epsilon(t)$ represent the vector of coalitions' excesses, formally given as:
\begin{eqnarray*}
\epsilon(t)=\left\{\epsilon_S(t)\right\}_{N\supseteq S\neq \emptyset}.
\end{eqnarray*}

We are interested in answering two main questions for this class of games. 
\begin{itemize} \item {\bf Question 1:} Are there allocation rules such that the average allocations converge? If yes, let us denote by 
$\mathcal{A}_0$ the set where the average  allocations converge to. Can we make it converge to 
the core of the average game $\mathcal{A}_0 \subseteq C(v_{nom})$? Can we guarantee the convergence to a specific 
point of the core, call it nominal allocation $a_{nom}$, that we have {\it a priori} selected? 
\item {\bf Question 2:} Are there allocation rules such that the coalitions' excesses $\epsilon(t)$ converge to an 
{\it a priori} given cone $\Sigma_0$, say for instance the nonnegative $m$-dimensional orthant $\mathbb R_+^{m}$, or any direction $\alpha t$ for $t\geq 0$ with fixed $\alpha\in \mathbb R_+^{m}$?  
\end{itemize}

To motivate the above questions think of a situation where the objective of the central planner is to maintain the stability of grand coalition in an average sense, while controlling the coalitions' excesses at each time $t \in \mathbb R_+$.

We are now in the position of providing a formal and generic statement of the problem. Henceforth, we use the symbol w.p.1 to mean ``with probability one''.
\begin{problem}
\label{prob:MP}
Find an allocation rule $f: \mathbb R^m \rightarrow \mathcal A\in \mathbb R^n$, such that if
{$a(t)=f\left(\epsilon(t)\right)$}
then i) $\lim_{t\rightarrow \infty} \bar a(t) \in \mathcal{A}_0\subseteq C(v_{nom})$ w.p.1, and ii) $\lim_{t\rightarrow \infty}\epsilon(t) \in \Sigma_0 \subseteq \mathbb R_+^{m}$ w.p.1.
\end{problem}
%
Observe that because of the random nature of the coalitions' values $v(t)$, both the excesses $\epsilon(t)$
and the allocations $a(t)$ are random and as such we look at the convergence of $\bar a(t)$ w.p.1. Essentially, we require that the probability of $\bar a(t)$ converging in the limit to $\mathcal{A}_0\subseteq C(v_{nom})$ is 1. Similarly for $\epsilon(t)$ and  $\Sigma_0$. This type of convergence is also known as \textit{almost sure} convergence \cite{LF96}.


We will show that if the planner has full observation of $\epsilon(t)$ at every time $t$ then 
the above problem is solvable even under the very strict condition of $\mathcal{A}_0=a_{nom}$ 
and $\Sigma_0=\alpha t$ $t\geq 0$ with fixed $\alpha$. 
Conversely, if the planner has partial observation of $\epsilon(t)$ in that he only knows the sign of each component of $\epsilon(t)$, then the problem is still solvable but under the relaxed condition of $\mathcal{A}_0 = C(v_{nom})$
and $\Sigma_0 \subseteq \mathbb R_+^{m}$.

\subsection{Motivations}\label{sec:motivations}
Dynamic coalitional games capture coordination in a number of network 
flow applications. Network flows model flow of goods, materials, or
other resources between different production/distribution sites~\cite{BBP10}. 
We next provide a supply chain application that justifies the model under study. 

\begin{figure}[ht]
\centering
\subfigure[Five trucks (cycles) leaving ${\bf v}_0$ and serving coalitions $\{ {\bf v}_1,\ldots,{\bf v}_4\}$, $\{ {\bf v}_5,\ldots,{\bf v}_7\}$,
$\{ {\bf v}_8\}$, $\{ {\bf v}_9\}$, and $\{ {\bf v}_{10}\}$ respectively.]{
\includegraphics[scale=.38]{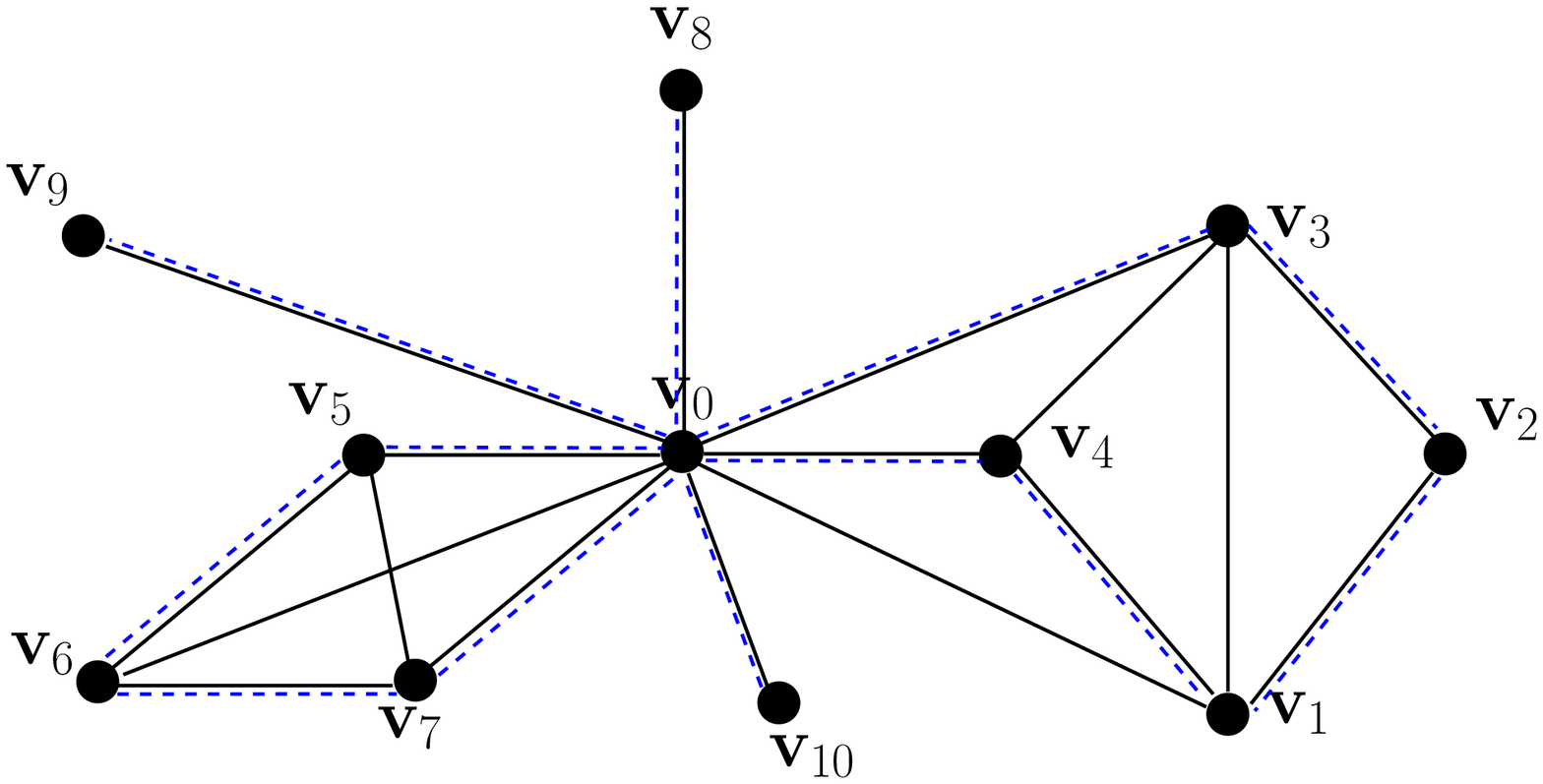}
\label{fig:subfig1-1}
}
\hskip .6cm
\subfigure[One single truck (cycle) leaving ${\bf v}_0$ and serving coalition $\{ {\bf v}_1,\ldots,{\bf v}_{10}\}$.]{
\includegraphics[scale=.38]{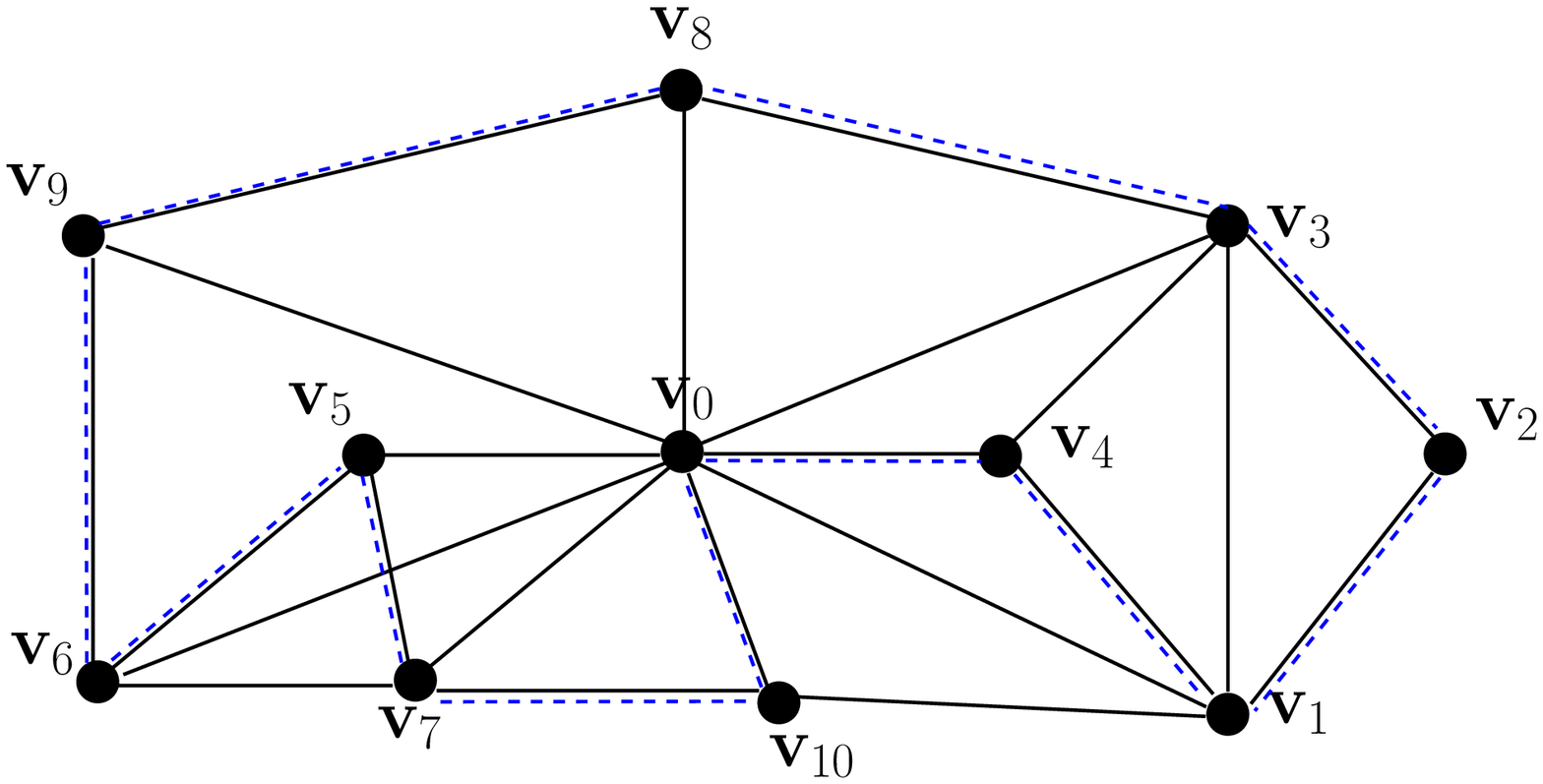}
\label{fig:subfig2-2}
}
\caption{Example of a distribution network}
\label{fig:subfigureExample1}
\end{figure}

A single warehouse ${\bf v}_0$ serves a number of retailers 
${\bf v}_i$, $i=1,\ldots,n$, each one facing a demand $d_i(t)$ 
unknown but bounded by pre-assigned values  $d_i^{\min} \in \mathbb{R}$ and
$d_i^{\max} \in \mathbb{R}$ at any time period $t\ge0$.    
After demand $d_i(t)$ has been realized, 
retailer ${\bf v}_i$ must choose to either fulfill the demand or not. 
The retailers do not hold any private inventory and, therefore, 
if they wish to fulfill their demands, they must
reorder goods from the central warehouse. Retailers benefit from joint reorders
as they may share the total transportation cost $K$ (this cost could also 
be time and/or players dependent). In particular, if retailer ${\bf v}_i$ 
``plays'' individually, the cost of reordering coincides with the full
transportation cost $K$. Actually, when necessary a single truck will serve 
only him and get back to the warehouse. This is illustrated by the dashed 
cycles $({\bf v}_0,{\bf v}_8,{\bf v}_0)$, $({\bf v}_0,{\bf v}_9,{\bf v}_0)$, and $({\bf v}_0,{\bf v}_{10},{\bf v}_0)$ in the network of 
Figure~\ref{fig:subfigureExample1}. The cost of not reordering is 
the cost of the unfulfilled demand $d_i(t)$. 

If two or more retailers ``play'' in a coalition, 
they agree on a joint decision (``everyone reorders'' or
``no one reorders''). The cost of reordering for the coalition
also equals the total transportation cost that must be
shared among the retailers. In this case, when necessary a single truck 
will serve all retailers in the coalition and get back to the warehouse. 
This is illustrated, with reference to coalition $\{{\bf v}_1,\ldots,{\bf v}_4\}$ 
by the dashed cycle $({\bf v}_0,{\bf v}_4,{\bf v}_1,{\bf v}_2,{\bf v}_3,{\bf v}_0)$ in
Figure~\ref{fig:subfig1-1}. A similar comment applies to the coalition 
$\{{\bf v}_5,{\bf v}_6,{\bf v}_7\}$ and the cycle 
$({\bf v}_0,{\bf v}_5,{\bf v}_6,{\bf v}_7,{\bf v}_0)$ in 
Figure~\ref{fig:subfig1-1}. The network topology in Figure~\ref{fig:subfig1-1}
describes the existing coalitions. This is clear  if we look at the subgraph induced by 
the vertex-set $\{{\bf v}_1,\ldots,{\bf v}_{10}\}$ (all vertices except ${\bf v}_0$) and observe that 
such a subgraph has 5 connected components, i.e., $\{ {\bf v}_1,\ldots,{\bf v}_4\}$, $\{ {\bf v}_5,\ldots,{\bf v}_7\}$,
$\{ {\bf v}_8\}$, $\{ {\bf v}_9\}$, and $\{ {\bf v}_{10}\}$ and that each component corresponds to an existing coalition.
The cost of not reordering is the sum of the unfulfilled 
demands of all retailers. How the players will share 
the cost is a part of the solution generated by the bargaining process.

Conversely, the subgraph induced  by $\{{\bf v}_1,\ldots,{\bf v}_{10}\}$ in  Figure~\ref{fig:subfig2-2} 
has a single connected component which means that all retailers ``play'' in the grand coalition and as such
one single truck (cycle) will leave ${\bf v}_0$ and serve all of them before returning to ${\bf v}_0$.
This is represented by the dashed cycle $({\bf v}_0,{\bf v}_4,\ldots,{\bf v}_{10})$ in the same figure.

The cost scheme can be captured by a game with the set 
$N=\{{\bf v}_1,\ldots,{\bf v}_n\}$ of players where 
the cost of a nonempty coalition $S\subseteq N$ is  given by
$$c_S(t)=\min\left\{K,\sum_{i\in S} d_i(t)\right\}.$$
Note that the bounds on the demand $d_i(t)$ reflect into the bounds 
on the cost as follows: for all nonempty $S\subseteq N$ and $t\ge0$, 
\begin{equation}
\label{eq:cs} \min\left\{\sum_{i\in S} K,d_i^{\min}\right\} 
\leq c_S(t)\leq \min\left\{K,\sum_{i\in S} d_i^{\max}\right\}.\end{equation}
To complete the derivation of the coalitions' values we need to compute the 
cost savings $v_S(t)$ of a coalition $S$ as the
difference between the sum of the costs of the coalitions of the
individual players in $S$ and the cost of the coalition itself, namely,
\[v_S(t) =\sum_{i\in S}c_{\{i\}}(t)-c_S(t).\]

Given the bound for $c_S(t)$ in~\eqref{eq:cs},
the value $v_S(t)$ is also bounded, as given: for any $S\subset N$ and $t\ge0$,
\begin{eqnarray*}
v_S(t) \leq \sum_{i\in S} \min\left\{K, d_i^{\max}\right\}
-\min\left\{K,\sum_{i\in S}d_i^{\min}\right\}.
\end{eqnarray*}
Thus, the cost savings (value) of each coalition is
bounded uniformly by a maximum value. 

Introducing time aspects into a static TU game opens the possibility for modeling aspects such as intertemporal transfers, patience and expectations of players/coalitions. A generic dynamic coalitional game description should capture these features.
In a repeated joint replenishment game as the one discussed above, allocation rules having the properties 
formalized in Problem  \ref{prob:MP}, encourage \textit{patient} retailers to ``play'' in the grand coalition, to coordinate their replenishment policies and therefore to 
reduce total transportation costs.  We say \textit{patient} retailers since condition i) in Problem  \ref{prob:MP} guarantees convergence to core on the long-run, i.e., in an average sense. Condition ii)  has the meaning of bounding the excesses during the transient (before convergence occurs).

\section{Flow transformation based dynamics}\label{sec:FT}
\begin{figure}[h]
\centering
\includegraphics[width=8cm]{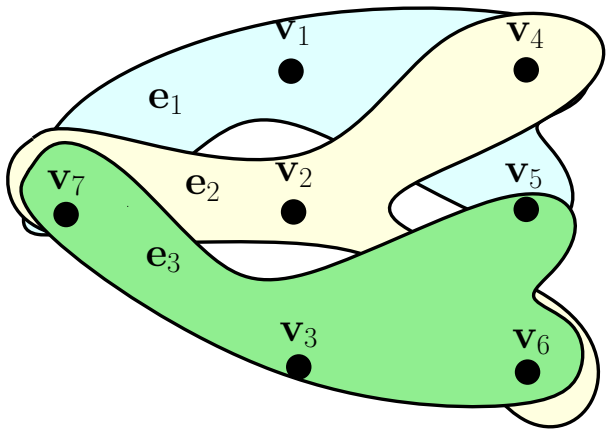}
\caption{Hypergraph $\mathcal{H}:=\{V,E\}$ for a $3$-player coalitional game.}
\label{fig:Hypergraph2}
\end{figure}

\begin{figure} [htb]
\centering
\def\svgwidth{0.5\columnwidth}
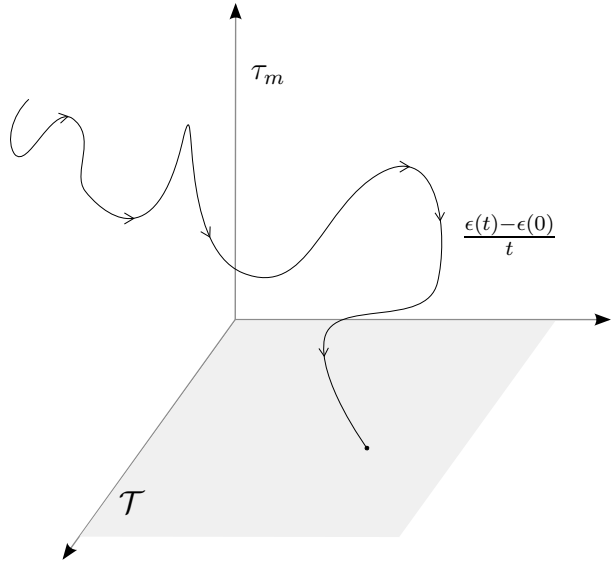
\caption{Trajectory for $\frac{\epsilon(t)-\epsilon(0)}{t}$.}  \label{fig:Traj}
\end{figure}

The basic idea of our solution approach is to recast the problem into a flow control one.
To do this, consider the hyper-graph $\mathcal H$ with vertex set $V$ and edge set $E$ as:
$$\mathcal H:=\{V,E\}, \quad V=\{ {\bf v}_1,\ldots,{\bf v}_m\}, \quad E:=\{{\bf e}_1,\ldots,{\bf e}_n\}.$$
{Figure \ref{fig:Hypergraph2} depicts an example of hypergraph for a 3-player coalitional game.} The vertex set $V$ has one vertex per each coalition whereas the edge set $E$ has one edge per each player.
A generic edge $i$ is incident to a vertex ${\bf v}_j$ if the player $i$ is in the coalition associated to ${\bf v}_j$.
So, incidence relations are described by matrix
$B_{\mathcal H}$
whose rows are the characteristic vectors $c^S\in \mathbb{R}^n$.
We recall that the components of a characteristic vector $c^S_i=1$ if $i\in S$
and $c^S_i=0$ if $i\notin S$.
The flow control reformulation arises naturally if we view allocation $a_i(t)$ as the flow on edge ${\bf e}_i$
and the coalition value $v_S(t)$ of a generic coalition $S$ as the demand in the corresponding vertex ${\bf v}_j$.
In view of this, allocation in the core translates into
over-satisfying the demand at the vertices. Specifically,
\begin{align}
\label{in} a(t) \in C(v(t)) \quad \Leftrightarrow \quad B_{\mathcal H}  a(t) \geq   v(t),
\end{align} with the  last inequality satisfied with the equal sign due to the efficiency condition of the core, i.e, $\sum_{i=1} a_i(t) = v_m(t)$, where $v_m(t)$ denotes the $m$th component of $v(t)$ and is equal to the grand coalition value $v_N(t)$.  Now, since $v(t)$ is unobservable by the planner at time $t$, we need to introduce some allocation error dynamics which accounts for the derivatives
of the excesses. Since $\epsilon(t)$ represents the coalition excess, we have:
\begin{equation}\label{epsdot}\dot \epsilon(t)= B_{\mathcal H}  a(t) -   v(t),\quad v(t)\in \mathcal V.\end{equation}
%
%
Note that the above differential equation
admits  a solution at least in the sense of Filippov \cite{F60}.
From (\ref{in}) and by averaging and taking the limit in (\ref{epsdot}), we can reformulate 
Problem \ref{prob:MP} as a flow control problem where a controller wishes to drive the 
quantity  $\lim_{t\rightarrow \infty}\frac{\epsilon(t)-\epsilon(0)}{t}$
to the target set  $\mathcal{T}$, defined below, w.p.1 { (see, e.g.,  Fig. \ref{fig:Traj})}: $$ \mathcal T:=\{\tau \in \mathbb R^m:\, \tau_m=0, \tau_j \geq 0, \, \forall j=1,\ldots, m-1\}.$$
Note, $\tau_m=0$ due to efficiency of allocations.\\

\begin{remark}
Driving the average allocations to a particular point
$a_{nom}\in\mathcal{A}_0\subseteq C(v_{nom})$ results in reaching a specific point in the target set $\mathcal{T}$. To see this, note that when $\lim_{t\rightarrow \infty}\bar{a}(t)=a_{nom}$ we have
$\mathcal{T}\ni B_\mathcal{H}a_{nom}-v_{nom}\geq 0$ due to the property of the core.
Thus, we also have that $\lim_{t\rightarrow \infty}\frac{\epsilon(t)-\epsilon(0)}{t}$
is driven to the point $B_\mathcal{H}a_{nom}-v_{nom} \in \mathcal{T}$.
\end{remark}
\noindent
The inequality condition in \eqref{in} is transformed into equality type by introducing, from standard LP techniques, $m-1$ surplus variables (one per each coalition other than the grand coalition). This increases the dimension of the control space of the planner from $m$ to $n+m-1$ and the dynamics \eqref{epsdot} can be
rewritten as follows:
\begin{eqnarray}\label{eq:convcond}
\dot{x}(t)=Bu(t)-v(t),~v(t)\in \mathcal{V}
\end{eqnarray}
where $B=\left[B_\mathcal{H}~ \Bigl| \begin{array}{c}-I_{m-1} \\ 0\end{array}\right] \in \mathbb{R}^{m\times n+m-1}$. Variable $x(t)$ represents the state of the system and captures deviation from the balanced system, i.e., the system characterized 
by $a_{nom}$ and $v_{nom}$. We introduce the set of feasible
controls as:
\begin{eqnarray}
U:=\left\{u(t)\in \mathbb{R}^{n+m-1} : u(t)=[a^T(t)~~ s^T(t)]^T,~ a(t)\in \mathcal A, ~s(t)\geq 0\right\}.
\label{eq:inpspace}
\end{eqnarray}
\noindent
Toward the reformulation of the problem as a stochastic stabilizability one, we introduce the following preliminary result.
%
%
%

\begin{lemma} 
\label{lemma:coreconv}
If the variable $x(t)$ is \textit{asymptotically stable almost surely}, i.e., (\ref{x=0}) holds true, then 
the average allocations converge to the core of the average game w.p.1. as expressed by (\ref{eq:longtermcore}), and 
the excesses converge to the cone  $\mathbb R_+^{m}$ w.p.1. as described in (\ref{eq:longtermcore111}):
\begin{eqnarray}
\label{x=0}\lim_{t \rightarrow \infty} {x}(t)=0, \quad \mbox{w.p.1.}\\
\label{eq:longtermcore}
\lim_{t \rightarrow \infty} \bar a(t) \in C(v_{nom}),\, \mbox{w.p.1}\\
\label{eq:longtermcore111} \lim_{t\rightarrow \infty}\epsilon(t) \in \mathbb R_+^{m},\, \mbox{w.p.1}.
\end{eqnarray}
\end{lemma}
\begin{proof} 
To see why (\ref{x=0}) implies (\ref{eq:longtermcore}), observe that if 
$\lim_{t \rightarrow \infty} {x}(t)=0$ w.p.1. then $\lim_{t\rightarrow \infty} \frac{x(t)-x(0)}{t}=0$ w.p.1.
and therefore, by integrating and dividing by $t$ in (\ref{eq:convcond}) 
also $\lim_{t\rightarrow \infty}B\bar{u}(t)-\bar v(t)=0$ w.p.1. The latter can be rewritten as $\lim_{t\rightarrow \infty} B\bar u(t)= v_{nom}$ w.p.1, and as from (\ref{eq:inpspace}) $\bar s(t)=B_\mathcal{H}\bar a(t)-\bar v(t) \geq 0$ and
$v_{nom}$  is balanced by Assumption 2 then we conclude that $\lim_{t\rightarrow \infty}\bar a(t)\in C(v_{nom})$ w.p.1.

To see why (\ref{x=0}) implies (\ref{eq:longtermcore111}), observe that if 
$\lim_{t \rightarrow \infty} {x}(t)=0$ w.p.1., from (\ref{eq:inpspace}) and under the assumption
$x(0)=\epsilon(0)=0$, then $\lim_{t\rightarrow \infty} \epsilon(t) = \lim_{t\rightarrow \infty} \tilde s(t) \geq 0$ and 
(\ref{eq:longtermcore111}) is proved. \end{proof}
\noindent
It is worth noting that condition (\ref{eq:longtermcore}) is part of Problem \ref{prob:MP}. In other words when solving Problem \ref{prob:MP} we always guarantee (\ref{eq:longtermcore}). If this is clear then, we can use the above lemma to rephrase Problem \ref{prob:MP}. In doing this we need to make a partial distinction between cases i) and ii). More specifically, case ii) where $\mathcal{A}_0=C(v_{nom})$ can be restated as follows: 
\begin{eqnarray}
\text{Find } u(t):=\phi(x(t))\in U \text{ such that } \lim_{t\rightarrow \infty}{x}(t)=0 \text{ w.p.1.}
\label{eq:coreconv}
\end{eqnarray}
%
%

%

Note that if we wish to reach a specific point $a_{nom}$ then the condition (\ref{eq:longtermcore}) is only necessary and the resulting problem  is a stricter version of (\ref{eq:coreconv}).
%
%
%

\section{Main results}\label{sec:mainresults}
In this section we present the three main results of this work. The first one relates to the case where the planner has full observation on $x(t)$ in which case the average allocation can be driven to a specific point in the Core of the average game. The second result applies to the case where the planner has partial observation on $x(t)$, and convergence to the Core can  still be guaranteed but not to a specific point of the Core. The third result highlights connections of the implemented solution approach to the approachability principle \cite{B56,L02} and attainability principle \cite{BLS12,LSB11}.  

\subsection{Full information case} 
\label{sec:full}
In this section, we solve Problem \ref{prob:MP} with $\mathcal{A}_0=a_{nom}$ and $\Sigma_0=\alpha t, t\geq 0$ with fixed $\alpha$ under the assumption that the planner has full observation of the excesses $\epsilon(t)$ and therefore $x(t)$ as well.
We recall that inferring $x(t)$ from $\epsilon(t)$ is possible as the surplus $s(t)$ is selected by the planner. As we have said before, the problem that we solve is a stricter version of (\ref{eq:coreconv}). This version derives from augmenting the state of dynamics (\ref{eq:convcond}) as explained in the rest of this section. Before introducing the augmentation technique let us  assume that  the fluctuations of the coalitions' values around the mean $v_{nom}$ are independent of the state $x(t)$. We formalize this in the next assumption where we denote by $\Delta v(t)=v(t)-v_{nom}$ the above fluctuations.
\begin{assumption}
\label{eq:ass5}
The state $x(t)$ and the coalitions' values fluctuations $\Delta v(t)$ are independent.
\end{assumption}
\noindent
%
Introducing the fluctuations $\Delta v(t)$ allows us to rewrite dynamics \eqref{eq:convcond} in a more convenient way. To do this, note first that, as $u(t)=[a(t)^T ~s(t)^T]^T$ and from $Bu_{nom}=v_{nom}$, if $a_{nom}$ is fixed
then $s_{nom}\in \mathbb R_+^{m-1}$ and therefore also  $u_{nom}=[a_{nom}^T \: s_{nom}^T]^T$ are fixed. Let us denote
$\Delta u(t)=u(t)-u_{nom}$. Dynamics \eqref{eq:convcond} can be
rewritten as follows:
\begin{eqnarray*}
\dot{x}(t)=Bu(t)-v(t)&=&Bu(t)-(v_{nom}+(v(t)-v_{nom}))=Bu(t)-v_{nom}-\Delta v(t)\\
&=&B\left(u(t)-u_{nom}\right)-\Delta v(t)
=B \Delta u(t)-\Delta v(t)
\end{eqnarray*}
We mentioned before that we will focus on a stricter version of (\ref{eq:coreconv}). 
We do this by augmenting the state as shown next. First, denote by $B^{\dagger}$ a generic pseudo inverse matrix of $B$ and complete matrices $B$ and $B^\dagger$ with matrices $C$ and $F$ such that
\\
\begin{equation}\label{inverse_rel}
\left[ \begin{array} {c}     B \\ C
\end{array} \right]
~ \left[ \begin{array} {cc}     B^\dagger & F
\end{array} \right]
= I.
\end{equation}
\\
Then, building upon the new square matrix $\left[ \begin{array} {c}     B \\ C
\end{array} \right]$, let us consider the augmented system
\begin{equation} \label{extended_system}
\begin{array} {rcl}
\dot x(t) & = & B \Delta u(t) - \Delta v(t) \\
\dot y(t) & = & C \Delta u(t).
\end{array}
\end{equation}
Here  we assume that $v(t)$ is independent of $y(t)$ as well.
After integrating the above system (see (\ref{nv}), right) we define a new variable $z(t)$ as follows:
\\
\begin{equation}\label{nv}
z(t) = \left[ \begin{array} {cc}     B^\dagger& F
\end{array} \right]~
\left[ \begin{array} {c}     x(t) \\ y(t)
\end{array} \right],~~~
\left[ \begin{array} {c}     x(t) \\ y(t)
\end{array}\right]
= \left[ \begin{array} {c}     B \\ C
\end{array} \right]~
z(t).
\end{equation}
It turns out that to drive $x(t)$ to zero w.p.1, and obtain $u_{nom}$ as average allocation on the long run,
we can rely on a simple function $\hat{\phi}(.)$, which depends on $z(t)$. Before introducing this function,
for future purposes observe that the dynamics for $z(t)$ satisfies the first-order differential equation:
\begin{figure} [htb]
\centering
\def\svgwidth{0.88\columnwidth}
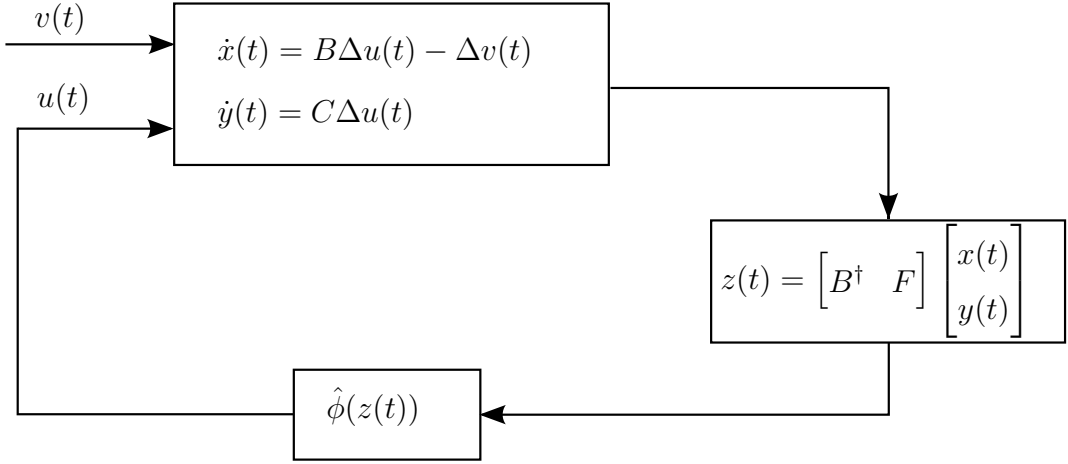
\caption{Dynamical System}  \label{fig:Block}
\end{figure}
\begin{equation} \label{new_system}
\begin{array}{lll}
\dot z(t) & = & \left[ \begin{array} {cc}     B^\dagger& F
\end{array} \right]~
\left[ \begin{array} {c}     \dot x(t) \\ \dot y(t)
\end{array} \right]\\
&=& \left[ \begin{array} {cc}     B^\dagger& F
\end{array} \right]~
\left[ \begin{array} {c}    B \\ C
\end{array} \right] \Delta u(t) - \left[ \begin{array} {cc}     B^\dagger& F
\end{array} \right] \left[ \begin{array} {c}    \Delta v(t) \\ 0
\end{array} \right] \\
 &=&\Delta u(t) - B^\dagger \Delta v(t).
\end{array}\end{equation}
%
%
Let $\Delta u^{min}$ and $\Delta u^{max}$ be the minimal and maximal values of $\Delta u(t)$ for the following constraints to hold true: $u(t)=u_{nom} + \Delta u(t) \in U$.
Then, let us formally define $\hat{\phi}(z(t))$ as:
\begin{equation}\label{cl}\hat \phi(z(t)) := u_{nom} + \Delta u(t) \in U, \quad  \Delta u(t)=sat_{[\Delta u^{min},~\Delta u^{max}]} (-z(t)),\end{equation}
\noindent
where with $sat_{[a,b]}(\xi)$  we denote the saturated function that, given a generic vector $\xi$ and lower and upper bounds $a$ and $b$
of same dimensions as $\xi$, returns
$$sat_{[a,b]}(\xi) \doteq \left \{
\begin{array}{lll}
b_i & \mbox{for all $i$} &  \xi_i > b_i\\
a_i & \mbox{for all $i$} &  \xi_i < a_i\\
\xi_i   & \mbox{for all $i$} &  a_i \leq   \xi_i \leq  b_i
\end{array}
\right. .$$
Now, taking the control $u(t)=\hat \phi(z(t))$, we obtain the dynamic system $\dot z(t) = B \hat \phi(z(t)) - v(t)$
as displayed in Fig. \ref{fig:Block}. 
With the above preamble in mind, we are ready to state the following convergence property.
\begin{theorem}\label{thm1}
Using the controller $\hat \phi(z(t))$, as in \eqref{cl}, we have 
$\lim_{t\rightarrow \infty} z(t)=0$ w.p.1 and therefore $\lim_{t \rightarrow \infty} \bar u(t)=u_{nom}$.
\end{theorem}

\noindent
In the next corollary, we use the previous result to provide an answer to Problem \ref{prob:MP}.
\begin{corollary}
\label{cor:cor1}
The state $x(t)$ is driven to zero w.p.1 as expressed in 
\eqref{eq:coreconv}, the average allocation converges to the nominal allocation 
i.e., $\lim_{t \rightarrow \infty} \bar a(t) = a_{nom}$, w.p.1 and the excesses converge to the direction
$\Sigma_0=\alpha t$ with $\alpha = s_{nom}$, i.e., $\lim_{t \rightarrow \infty} \epsilon(t) \in \Sigma_0$. 
\end{corollary}
\begin{proof}
This is a direct consequence of the result proved in the previous theorem. From \eqref{nv}, and $[B^\dagger~ F]$ being a non singular matrix, we have $\lim_{t \rightarrow \infty} x(t)=0$ w.p.1. 
From the previous theorem we also have $\lim_{t \rightarrow \infty} \bar u(t)=u_{nom}$. Since $u(t)=[a^T(t)~s^T(t)]^T$, we have that
$\lim_{t \rightarrow \infty} \bar a(t)=a_{nom}$ and $\lim_{t \rightarrow \infty} \epsilon(t) = \tilde s(t) = s_{nom} t$.
 \end{proof}

\noindent
To summarize, in the full information case, the controller $u(t)$ defined by \eqref{cl} induces an allocation sequence $a(t)$ such that the average $\bar{a}(t)$ converges to $\mathcal{A}_0=a_{nom}$ and the excesses approach 
$s_{nom} t$.

\subsection{Partial information case}
\label{sec:partial}
In the previous section we observed that if the planner has full observation of the excesses and therefore of $x(t)$ then he can design an allocation rule so that the average allocations are driven to
$a_{nom}$ and the excesses approach $s_{nom}t$. In this section, we solve Problem \ref{prob:MP} with $\mathcal{A}_0 = C(v_{nom})$ and under the assumption that the planner has partial observation of $x(t)$. In particular, we assume that the planner observes the sign of $x(t)$ for all 
$t \in \mathbb R_+$. 
An information structure based on the sign of $x(t)$ has an oracle-based interpretation which we discuss in detail in Subsection \ref{sec:obi}.
%

\begin{figure}[h]
\centering
\def\svgwidth{0.7\columnwidth}
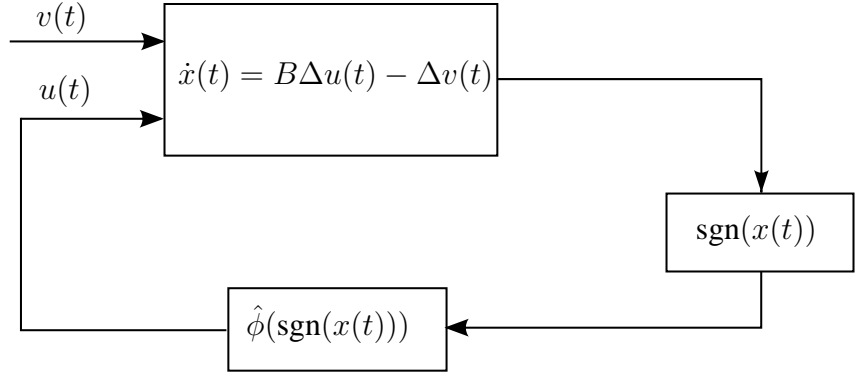
\caption{Dynamical System}  \label{fig:Blockpartial}
\end{figure}

Similarly to the previous section, suppose that we know a particular allocation
$a_{nom}$ in the core $C(v_{nom})$, and let us study the
convergence properties of the average allocations. In particular, using an allocation rule $u(t)=\phi(x(t))$, we require that $x(t)$ 
satisfying the dynamics $\dot x(t) = B \phi(x(t))- v(t)$, converge to zero in probability.
%
In this section, we state the second main result of this work which provides a solution to Problem \ref{prob:MP}
with partial information structure.
To do this, let us denote again by $B^{\dagger}$ a generic pseudo inverse matrix of $B$ and take a feasible allocation $u_{nom}$ such that $$B u_{nom}= v_{nom}:=\lim_{t\rightarrow \infty} \bar v(t), \quad u_{nom}\in U.$$ Also, for future purposes, define
a function $\hat{\phi}(.)$, which depends only on the sign of $x(t)$, as follows:
\begin{equation}\label{c2}\hat \phi(sgn(x(t))) := u_{nom} + \Delta u(t) \in U, \quad  \Delta u(t)=- \delta B^{\dagger} sgn(x(t)).\end{equation} Now, taking the control $u(t)=\hat \phi(sgn(x(t)))$, we obtain the dynamic system $\dot x(t) = B \hat \phi(sgn(x(t))) - v(t)$
as displayed in Fig. \ref{fig:Blockpartial}. 
Now, we state the following convergence property.
\begin{theorem}\label{thm2}
Using the controller $u(t)=\hat \phi(sgn(x(t)))$ as in \eqref{c2} we have 
$\lim_{t\rightarrow \infty} x(t)=0$ w.p.1.
\end{theorem}


\begin{corollary}
\label{cor:cor2}
The average allocation converges to the core of the average game as in (\ref{eq:longtermcore}) and the excesses $\epsilon(t)$ converge to $\mathbb R_+^{m}$ as in (\ref{eq:longtermcore111}).
\end{corollary}
\begin{proof}
Direct consequence of Theorem \ref{thm2} and Lemma \ref{lemma:coreconv}. 
\end{proof}

\subsubsection{Oracle-based interpretation}\label{sec:obi}


In this subsection we elaborate more on the partial information structure. 
In particular, we highlight how the feedback on state $x(t)$ can be reviewed as the result of an oracle-based procedure.
To see this, assume that the planner knows the sign of $x(t)$. Since $x(t)=\left(\epsilon(t)-\tilde{s}(t)\right)-\left(\epsilon(0)-x(0)\right)$, $\text{sgn}(x(t))$ reflects over-satisfaction of coalitions with respect to the threshold $\tilde s(t)$. In particular, take without loss of generality $\epsilon(0),x(0)=0$, then
with reference to component $j$, the sign of $x_j(t)$ yields:
\begin{equation}\label{thr}sgn\left(x_j(t)\right):=\left\{\begin{array}{cl} 1 & \epsilon_j(t) > \tilde s_j(t)\\
0 & \epsilon_j(t) = \tilde s_j(t)\\
-1 & \epsilon_j(t) < \tilde s_j(t).\end{array}\right.\end{equation}
To summarize, we can think of a
situation where the planner approaches an oracle that tells him the sign of $x(t)$. Since $s(t)$ is chosen by the planner for every $t$, the accumulated surplus, $\tilde{s}(t)$, is given as an input to the oracle. The oracle  returns ``yes'' 
if the actual  excess is greater than $\tilde{s}(t)$ and ``no'' otherwise. The use of an oracle is an element in common with the ellipsoid method in optimization and with a large literature \cite{WN88} on cutting planes.

Recall that nonnegativeness of the threshold has its roots in the feasibility condition $u(t)\in U$ for all $t \geq 0$
with feasible set $U$ as in \eqref{eq:inpspace}.

Nonnegativeness of the threshold provides us with a further comment on the information available to the planner.
Actually, from the first condition in (\ref{thr}), we can conclude that coalitions associated to a positive state $x(t)$
are certainly in excess. This is clear if we observe that $sgn\left(x_j(t)\right)=1$ implies
$\epsilon_j(t) > \tilde s_j(t) \geq 0$.
We can then summarize the information content available to the planner as follows, where $S$ is the generic coalition  associated with component $j$:
$$sgn\left(x_j(t)\right):=\left\{\begin{array}{cl} 1 & \mbox{then coalition $S$ in excess}\\
 -1,0 & \mbox{nothing can be said}.\end{array}\right.$$

Trivially, the development in the full information case in Section \ref{sec:full}, which is all based on control strategy (\ref{cl}), fits the case where $x(t)$ is revealed completely.
In this last case, the fact that the planner knows $x(t)$ implies 
that he knows $\epsilon(t)$ as well. Also, it is intuitive to infer that in this last
set up, exact knowledge of $x(t)$ can only influence positively the planner in terms
of speed of convergence of allocations to the core of the average game.
%
\begin{remark}
As the planner knows a priori the nominal game and a corresponding
nominal allocation vector, a natural question that arises is
why one has to design an allocation rule as given
by (\ref{cl}) and (\ref{c2}) instead of a stationary rule $\hat{\phi}(.)=u_{nom}$.
The rules given by (\ref{cl}) and (\ref{c2}) intuitively translate to meeting the demands of coalitions in an average sense. This feature reflects patience aspect of coalitions in a dynamic setting, i.e., even if a demand is not met instantaneously a coalition is willing to wait
and stay in the grand coalition as the demand is fulfilled in an average sense. 
\end{remark}
\subsection{Connections to Approachability and Attainability}
\subsubsection{Approachability}
Approachability theory was developed by Blackwell in 1956 \cite{B56}
and is captured in the well known Blackwell's Theorem. 
Along the lines of Section 3.2 in \cite{L02}, we recall next the geometric (approachability) principle
that lies behind Blackwell's Theorem. The goal of this section is to show 
that such a geometric principle shares striking similarities with the solution approach used in the previous sections. 

To introduce the approachability principle, let $\Phi$ be a closed and convex set in $\mathbb R^m$ and let $P(y)$ be the projection of any point $y \in \mathbb R^m$ (closest point to $y$ in $\Phi$). Also denote by $\bar y_k$ the average of $y_1\ldots,y_k$, i.e., $\bar y_k= \frac{\sum_{t=0}^k y_t}{k}$ and let $dist(\bar y_k,\Phi)$ be the euclidean distance between point $\bar y_k$ and set $\Phi$. 


\begin{lemma}\label{app_princ}(Approachability principle \cite{L02})
Suppose that a sequence of uniformly bounded vectors $y_k$ in $\mathbb R^m$ satisfies condition (\ref{appr}),
\begin{equation}\label{appr} [\bar y_k - P(\bar y_k)]^T [y_{k+1}-P(\bar y_k)] \leq 0, \end{equation}
then $\lim_{k \rightarrow \infty} dist(\bar y_k,\Phi)=0$.
\end{lemma}

Now, to make use of the above principle in our set up, let us consider the discrete time analog of the excess dynamics (\ref{eq:convcond}):
$$x_{k+1}=x_k+B\Delta u_k-\Delta v_k,$$ and define a new variable $y_k=x_{k}-x_{k-1}$ so that we can look at the sequence of $y_k$ in $\mathbb R^m$. 
Likewise, consider the discrete time version of control (\ref{c2}) as displayed below: 
\begin{equation}\label{c2d}
\hat \phi(sgn(x_k)) := u_{nom} + \Delta u_k \in U, \quad  \Delta u_k=- \delta B^{\dagger} sgn(x_k-x_0).\end{equation} 
We are now in a position to state the main result of this section.
\begin{theorem} 
\label{thm:approach}
Using the controller $u_k=\hat \phi(sgn(x_k-x_0))$ as in \eqref{c2d} we have  that
\begin{itemize}
\item [i)] the vector $0$ is approachable by the sequence $\bar y_k$,
\begin{equation}
\label{eq:longtermcorediscr1}\lim_{k \rightarrow \infty} \bar  y_k = 0,\,\mbox{w.p.1},\end{equation} and therefore
\item [ii)] the average allocations converge to the core of the average game, 
\begin{equation}
\label{eq:longtermcorediscr}
\lim_{k \rightarrow \infty} \bar a_k \in C(v_{nom}),\, \mbox{w.p.1}.
\end{equation}
\end{itemize}
\end{theorem}
The strength of the above result is in that it sheds light on how the convergence problem dealt with in this work has a stochastic stability interpretation as well 
as an approachability one. 

\begin{remark}(Continuous-time approachability)
We can reformulate Theorem \ref{thm:approach} in the continuous time. To see this, 
let us first define $y(t):= \dot x(t)$. Next we need to derive the continuous time version of (\ref{appr}).
To this aim, let $t\rightarrow r(t)$ be a differentiable continuous time variable and let
$z(t)=\frac{r(t)-r(0)}{t}$, so $t\dot{z}(t)+z(t)=\dot{r}(t)$. Discrete time versions
are given as $z_k=\frac{1}{k}r_k$ and $z_{k+1}=\frac{1}{k+1}r_{k+1}$.
The approachability principle is given as
\begin{eqnarray*}
\left[z_k-P(z_k)\right]^T\left[\phi-P(z_k)\right]\leq 0
\end{eqnarray*}
where $\phi=(k+1)z_{k+1}-kz_k$. 
In continuous time the above condition translates to
\begin{eqnarray*}
\left[z(t)-P(z(t))\right]^T\left[\phi-P(z(t))\right]\leq 0
\end{eqnarray*}
and $\phi=(t+\Delta t)z(t+\Delta t)-t z(t)=t\left(z(t+\Delta t)-z(t)\right)+\Delta t z(t+\Delta t)$. We see that  $\frac{\phi}{\Delta t}= t\frac{z(t+\Delta t)-z(t)}{\Delta t}+z(t+\Delta t)$. Further, as $\Delta t \rightarrow 0$ we have $\lim_{\Delta t \rightarrow 0} \frac{\phi}{\Delta t}=t \dot{z}(t)+z(t) =\dot{r}(t)$. The approachability principle in continuous time can then be reproposed  as
\begin{eqnarray}\label{ctapp}
\left[z(t)-P(z(t))\right]^T\left[\dot{r}(t)-P(z(t))\right]\leq 0,
\end{eqnarray}
which constitutes the continuous time version of (\ref{appr}).
%
%
%
%
If $\Phi=\{0\}$ we have $P(z(t))=0$ and $z^T(t)\dot{r}(t)\leq 0$. Now, taking $r(t)=x(t)$ we see that $z(t)$ is the average of $y(t)$. Then condition (\ref{ctapp}) guarantees that $z(t)$ converges to zero as well as $\bar y(t)$. But this implies that $\lim_{t \rightarrow \infty} \frac{x(t)-x(0)}{t}=0$
and therefore from Lemma \ref{lemma:coreconv} we arrive at (\ref{eq:longtermcore}) which represents the continuous time version of (\ref{eq:longtermcorediscr}).
\end{remark}

\subsubsection{Attainability}
Attainability is a new notion developed in \cite{BLS12,LSB11} in the context of 2-player continuous-time repeated games 
with vector payoffs. Attainability finds its roots in transportation networks, distribution networks, production networks applications. 
The main question is the following one: ``Under what conditions a strategy for  player 1 exists such that the cumulative payoff converges (in the lim sup sense) to a pre-assigned set (in the space of vector payoffs) independently of the strategy used by player 2''. 

Attainability shares similarities with two main notions in robust control theory \cite{B99}. 
The first notion is called \textit{robust global attractiveness} and refers to
the property of a set to ``attract'' the state of the system under a proper control strategy and
independently of the effects of the disturbance. The second notion is referred to as \textit{robustly
controlled invariance} and describes the property of a set to bound the state trajectory under
a proper control strategy and independently of the effects of the disturbance.
Both notions are used in the following formalization of the attainability principle. 
The principle is accompanied by a sketch of the proof but no formal proof is reported as attainability is the main focus of 
another paper and here it is just auxiliary to the solution of our main problem and also because 
the aforementioned two notions are well known in robust control theory. We 
refer the readers to \cite{B99} and \cite{BLS12,LSB11} for further details.  

Let $\Phi$ be a closed and convex set in $\mathbb R^m$ and 
consider a differentiable continuous-time variable $t \mapsto y(t)$ taking value in $\mathbb R^m$ for all $t \geq 0$. 
%
%
\begin{lemma}\label{att_princ}(Attainability principle \cite{BLS12,LSB11})
Suppose that the differentiable continuous-time variable $t \mapsto y(t)$ satisfies conditions (\ref{att1})-(\ref{att2}), 
\begin{eqnarray}\label{att1} [y(t) - P(y(t))]^T \left[\dot y(t)-P(y(t))\right]  < 0, &  y(t) \not \in  \Phi\\
\label{att2} n^T_{y(t)}\left[\dot y(t)-P(y(t))\right]  \leq 0, &  y(t) \in \partial \Phi,\end{eqnarray}
then $\lim_{t \rightarrow \infty} dist(y(t),\Phi)=0$.
\end{lemma}

Essentially, condition (\ref{att2}) is strictly related to the \textit{subtangentiality conditions} as formulated by Nagumo in 1942 and 
surveyed in \cite{B99}.  Such conditions are proven to characterize robustly controlled invariant sets. 
We provide a geometric perspective on such a condition  in Fig. \ref{fig:subfig2}. 
Consider a 2 player continuous-time repeated game and let $y(t)$ be the cumulative payoff up to time $t$. 
Denote by $Y$ the set of possible instantaneous vector payoffs, call them $\dot y(t)$, for a fixed strategy of player 1 and for varying strategy of player 2. Condition (\ref{att2}) is equivalent to $Y \subset H^-:= \{y\in \mathbb R^m| \, n_{y(t)} \dot y(t)\leq 0 \}$ and guarantees
that the cumulative payoff up to time $t+dt$ ($dt$ is the infinitesimal time interval) $y(t+dt)$ does not quit $\Phi$. 

As regards condition (\ref{att1}), suppose without loss of generality that 
$\Phi :=\{x\in \mathbb R^m| \,  V(x) \leq \hat \kappa\}$ for a fixed scalar $\kappa$.
Condition (\ref{att1}) establishes that the set $\Phi=\{x\in \mathbb R^m| \,  V(x) \leq \hat \kappa\}$  for any scalar $\hat \kappa$ satisfying 
$\hat \kappa > \kappa$ is a contractive set. By contractive set we mean that it is invariant and, whenever the state is on the
boundary, the control can ``push it towards the interior''. This is illustrated in Fig. \ref{fig:subfig1}. 
Let $Y$ and $y(t)$ have the same meaning as before. Condition (\ref{att1}) establishes that
$Y \subset H^-:= \{y\in \mathbb R^m| \, [y(t) - P(y(t))]^T \dot y(t)  < 0\}$ which implies that $dist(y(t+dt),\Phi) < dist(y(t),\Phi)$
and therefore $\Phi$ is robustly attractive.  

\begin{figure}[ht]
\centering
\subfigure[Robust global attractiveness: condition (\ref{att1}).]{
\includegraphics[scale=.38]{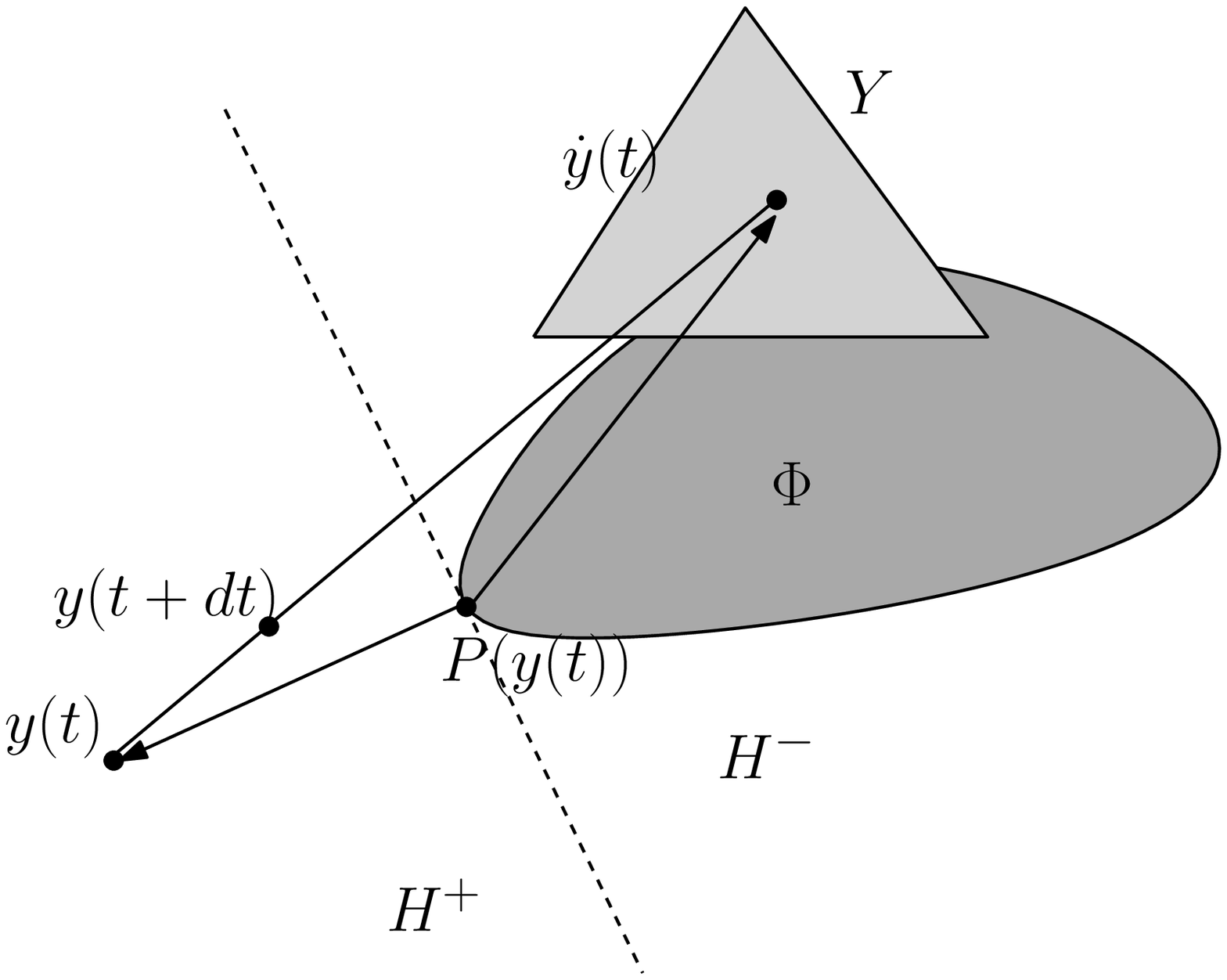}
\label{fig:subfig1}
}
\hskip .6cm
\subfigure[Robust control invariance: condition (\ref{att2}).]{
\includegraphics[scale=.38]{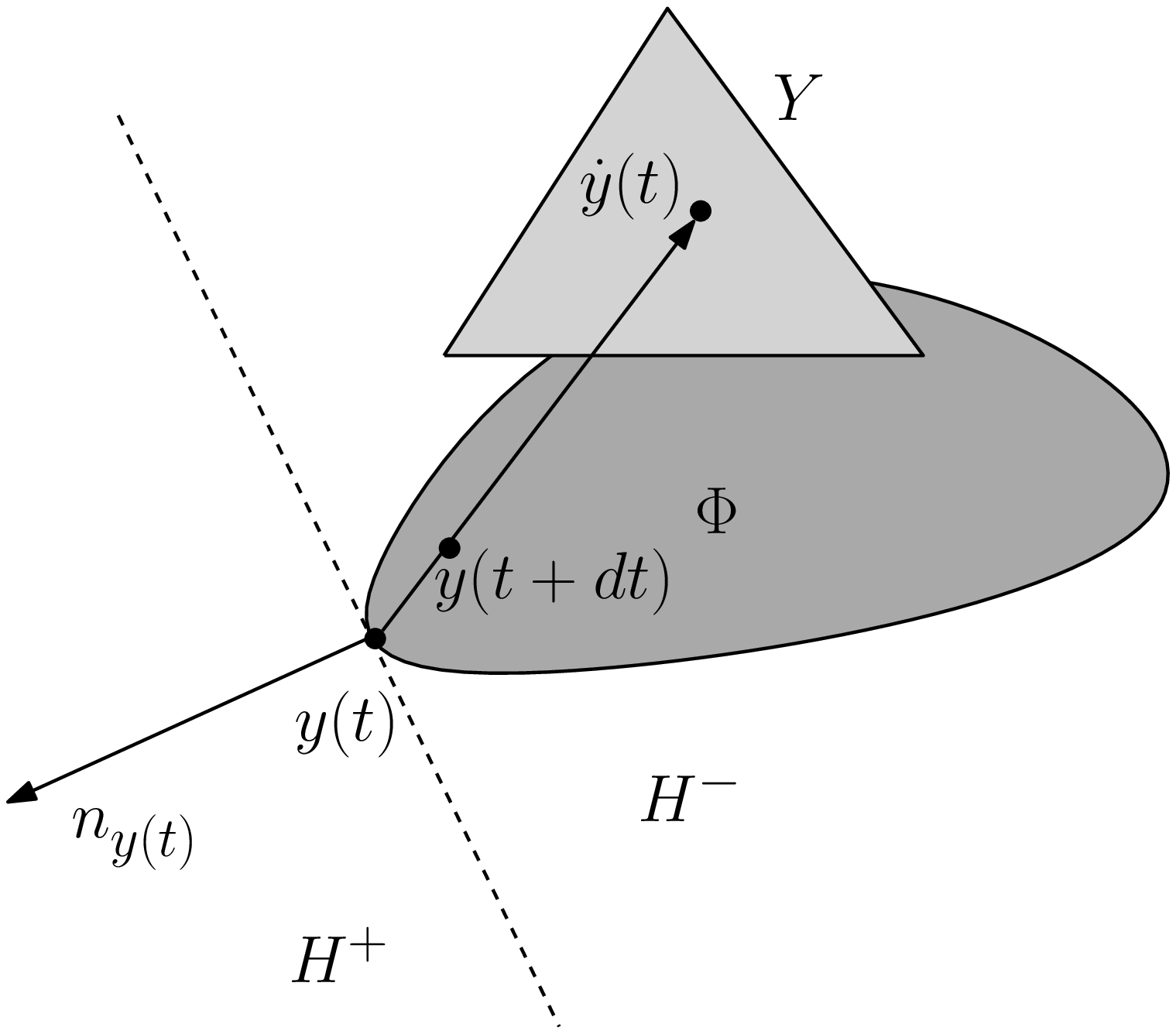}
\label{fig:subfig2}
}
\caption{Geometric representation of conditions (\ref{att1}) and (\ref{att2}).}
\end{figure}

Based on the above lemma, we can rephrase Theorem \ref{thm2} as follows.
\begin{theorem} 
\label{thm:attain}
Using the controller $u(t)=\hat \phi(sgn(x(t)))$ as in \eqref{c2} we have 
that the vector $0$ is attainable by $x(t)$.
\end{theorem}


\section{Derivation of the main results}\label{sec:resultsderiv}
\subsection{Proof of Theorem \ref{thm1}}
This proof is derived in the context of Lyapunov stochastic stability theory \cite{LF96}. We start by observing that using $u(t)=\hat \phi(z(t))$ we have:
\begin{equation}\label{eq:ds}\dot z(t) = B \hat \phi(z(t)) - v(t).\end{equation}
Consider a candidate Lyapunov function $V(z(t))=\frac{1}{2} z^T(t)z(t)$. The idea is to show that
$\mathbb E[\dot V(z(t))] < 0$ \footnote{Stochastic stability involves time derivative of the expectation of $V(x(t))$. However, since $V(.)$ is non-negative and smooth, the limit and expectation can be interchanged by using the dominated convergence theorem \cite{Williams}.} for all $t\geq 0$. Actually, the theory establishes that if the last condition holds true, then $V(z(t))$ is a 
supermartingale  and therefore by the martingale convergence theorem $\lim_{t \rightarrow \infty} V(z(t))=0$ w.p.1 (almost surely).
To see that $\mathbb E[\dot V(z(t))] < 0$ is true, observe that from (\ref{new_system}) we have
\begin{eqnarray*}
\mathbb E[\dot V(z(t))] & = &\mathbb E[z^T(t) \dot z(t)] \\
& = & \mathbb E[z^T(t) \Delta u(t)] - \mathbb E[z^T(t) B^\dagger \Delta v(t)] \\
&=&
\mathbb E[z^T(t) sat(-z(t))] < 0,
\end{eqnarray*}
where condition $\mathbb E[z^T(t) B^\dagger \Delta v(t)]=0$ is a direct consequence \footnote{If $\Delta v(t)$ is independent of $x(t)$ and $y(t)$ then $C \Delta v(t)$ is independent of $z(t)=Ax(t)+By(t)$.} of the assumption that $\Delta v(t)$ is independent of
$x(t)$ and $y(t)$. But the above condition implies that $\lim_{t \rightarrow \infty} V(z(t))=0$ w.p.1  and therefore also $\lim_{t \rightarrow \infty} z(t)=0$ w.p.1.
So far we have proved the first part of the statement, i.e., that the dynamic system \eqref{eq:ds} converges to zero w.p.1.
For the second part, after integrating dynamics (\ref{new_system}), we have
$$\lim_{t \rightarrow \infty} \frac{\int_0^t [\Delta u(\tau)-B^\dagger \Delta v(\tau)] d\tau}{t}=\lim_{t \rightarrow \infty}\frac{z(t)-z(0)}{t}=0.$$
This last condition together with the assumption $v_{nom}:=\lim_{t \rightarrow \infty} \bar v(t)$ yields
$$\lim_{t \rightarrow \infty} \frac{\int_0^t B^\dagger \Delta v(\tau) d\tau}{t}=\lim_{t \rightarrow \infty} \frac{\int_0^t \Delta u(\tau) d\tau}{t}=0$$
from which we can conclude $\lim_{t \rightarrow \infty} \bar u(t)=\lim_{t \rightarrow \infty} \frac{\int_0^t u_{nom} + \Delta u(\tau) d\tau}{t}=u_{nom}$ as claimed in the statement. 

\subsection{Proof of Theorem \ref{thm2}}
Consider a candidate Lyapunov function $V(x(t))=\frac{1}{2} x^T(t)x(t)$. The idea is to show that
$\mathbb E[\dot V(x(t))] < 0$ for all $t\geq 0$. For this to be true, it must be
\begin{eqnarray*}
\mathbb E[\dot V(x(t))] & = &\mathbb E[x^T(t) \dot x(t)] \\
& = & \mathbb E[x^T(t) B u(t)] - \mathbb E[x^T(t) v(t)] \\
&= &\mathbb E[x^T(t) B u_{nom}] + \mathbb E[x^T(t) B \Delta u(t)]- \mathbb E[x^T(t) v_{nom}]-\underbrace{\mathbb E[x^T(t) \Delta v(t)]}_{=0}\\
&=&\mathbb E[x^T(t) B \Delta u(t)] < 0.
\end{eqnarray*}
where condition $\mathbb E[x^T(t) \Delta v(t)]=0$ is a direct consequence of Assumption \ref{eq:ass5}.
But the above condition $\mathbb E[x^T(t) B \Delta u(t)] < 0$ is satisfied since $B \Delta u(t)= - \delta sgn(x)$, which in turn implies
$$\mathbb E[x^T(t) B \Delta u(t)] = \mathbb E[- \delta \|x(t)\|_1 ] < 0.$$
Then we obtain that $\lim_{t \rightarrow \infty} V(x(t))=0$ w.p.1  and therefore also $\lim_{t \rightarrow \infty} x(t)=0$ w.p.1 and this concludes the proof.

\subsection{Proof of Theorem \ref{thm:approach}}
We first prove that  (\ref{eq:longtermcorediscr1}) implies (\ref{eq:longtermcorediscr}). 
Invoking the discrete time reformulation of Lemma \ref{lemma:coreconv}, 
we can infer that 
$\lim_{k \rightarrow \infty} \frac{x_k - x_0}{k}=0$ w.p.1. 
implies $\lim_{k \rightarrow \infty} \bar a_k \in C(v_{nom}),\, \mbox{w.p.1}$.
Observing that $\bar y_k =\frac{x_k - x_0}{k}$ then we can conclude that $\lim_{k \rightarrow \infty} \bar y_k=0$ w.p.1
implies $\lim_{k \rightarrow \infty} \bar a_k \in C(v_{nom}),\, \mbox{w.p.1}$.



We now prove that using the controller $u_k=\hat \phi(sgn(x_k))$ as in \eqref{c2d}
then (\ref{eq:longtermcorediscr1}) holds true. To see this, let us
invoke the approachability principle in Lemma \ref{app_princ} and observe that a sufficient condition for approachability of $\bar{y}_k$ to 
$0$ is $\bar{y}^T_k y_{k+1}\leq 0$ for all $k$. This is evident if we take set $\Phi$ including only the zero vector, $\Phi=\{0\}$, and thus $P(\bar y_k)=0$ in (\ref{appr}). 
For the present case, using the definition of $y_k$, condition $\bar{y}^T_k y_{k+1}\leq 0$ would be
$\frac{1}{k}\left(x_k-x_0\right)^T\left(x_{k+1}-x_k\right)\leq 0$,
which implies $(x_k-x_0)^T B\Delta u_k - (x_k-x_0)^T \Delta v_k\leq 0$ for all $k$. Taking the expectation, 
from Assumption \ref{eq:ass5} we know that $\mathbb E[(x_k-x_0)^T \Delta v_k]=0$ and so we can write 
\begin{eqnarray*}\mathbb E[(x_k-x_0)^T B\Delta u_k - (x_k-x_0)^T \Delta v_k] & = &
\mathbb E[(x_k-x_0)^T B\Delta u_k] \\ & = & \mathbb E[(x_k-x_0)^T B   (- \delta B^{\dagger} sgn(x_k-x_0))] \leq 0.\end{eqnarray*}
From the above condition we derive that $\bar{y}^T_k y_{k+1}\leq 0$ w.p.1 for all $k$ and this concludes our proof.

\subsection{Proof of Theorem \ref{thm:attain}}
Let us invoke the attainability principle in Lemma \ref{att_princ} and observe that a sufficient condition for $x(t)$ to attain  
$0$ w.p.1 is that 
\begin{eqnarray}\label{att1-0} \mathbb E [x^T(t) \dot x(t)]  < 0, &  x(t) \not = 0 \\
\label{att2-0} \mathbb E [\dot x(t)] = 0, &  x(t) =0. \end{eqnarray}
This is evident if we take set $\Phi$ including only the zero vector, $\Phi=\{0\}$, and thus $P(x(t))=0$ in~(\ref{att1}) and~(\ref{att2}). 
Now, observe that condition (\ref{att1-0}) is equivalent to condition $\mathbb E [\dot V] < 0$ used in the proof of Theorem \ref{thm2}.
Condition (\ref{att2-0}) is also satisfied as $sgn(0)=0$ and this concludes our proof.

\section{Numerical illustrations}\label{sec:sim}

Consider a $3$ player coalitional TU game, so $m=7$, with values of coalitions in the following intervals:
\begin{eqnarray*}
&&v(\{1\})\in[0,4],~v(\{2\})\in[0,4],~v(\{3\})\in[0,4],\\
&&v(\{1,2\})\in[0,4],~v(\{1,3\})\in[0,6],\\
&&v(\{2,3\})\in[0,7],~ v(\{1,2,3\})\in[0,12].
\end{eqnarray*}
The convex set $\mathcal{V}$ is then a hyperbox characterized by the above intervals.
From Assumption \ref{ass:info}, the planner knows the long run average game, i.e., $\lim_{t\rightarrow \infty}\bar{v}(t)=v_{nom}$. Without loss of
generality we take the balanced nominal game be as $v_{nom}=[1~ 2~ 3~ 4~ 5~ 6~ 10]^T$. 
In other words, during the simulations we randomize the instantaneous games $v(t)\in \mathcal{V}$ so that it satisfies the average behavior given by:
\begin{eqnarray}
\lim_{t\rightarrow \infty}\frac{1}{t}\int_0^t v(\tau)d\tau=v_{nom}.
\label{eq:ingame}
\end{eqnarray}
Next, we describe an algorithm to generate $\mathbb P \in \Delta (\mathcal V)$ and therefore $v(t)\in \mathcal{V}$ such that 
the above condition holds true.

\begin{table}\normalsize
\begin{center}
\begin{tabular}{p{12cm}}\\
\toprule
\textbf{Algorithm} \\
\midrule
\textbf{Input:} Set $\mathcal{V}$ and value $v_{nom}$.  \\
\textbf{Output:} Probability function $\mathbb P \in \Delta (\mathcal V)$ to generate $v(t)\in \mathcal{V}$.\\
$\quad 1: $ \textbf{Initialize} Generate $m$ random points, $r_i\in \mathcal{V}\subset\mathbb{R}^m,~i=1,2,\cdots,m$,\\ 
$\quad 2: $ Solve $R.p=v_{nom}$, with $R=[r_1,~r_2,\cdots ~r_m]$,\\ 
$\quad 3: $ \textbf{If} $p\geq 0$ and ${\mathbf{1}^Tp}>0$, \textbf{then} go to ({\bf{4}}) else go to ({\bf{1}}),\\ 
$\quad 4: $ Rescale $R$ as $R=\left(\mathbf{1}^Tp\right)R$ and $p$ as $p=\frac{p}{\left(\mathbf{1}^Tp\right)}$,\\ 
$\quad 5: $ \textbf{If} $r_i\in \mathcal{V},~i=1,2,\cdots,m$, \textbf{then} go to ({\bf{6}}) else go to ({\bf{1}}).\\ 
$\quad 6: $ \textbf{STOP}\\ 
\bottomrule
\end{tabular}
\end{center}
\end{table}

By construction, $v_{nom}$ is in the relative interior of the convex hull generated by the columns of the matrix $R$. If an instance of the game $v(t)$ is chosen as $r_i$ with probability $p_i$ from the pair $(R,p)$, Assumption \ref{ass:info} is satisfied. For simulations we ran the algorithm $10$ times to generate $10$ $(R,p)$ pairs in $\mathcal{V}$. Further, from each pair $(R,p)$ we take $100,000$ random selections (using Matlab {\texttt{randsrc}} function) to realize $v(t)$. The step size is set to $\Delta=0.05$. The results
are averaged over the $10$ pairs. The nominal choice of allocations and surplus is taken as $u_{\text{nom}}=[2.5~ 3~ 4.5~ 1.5~ 1~ 1.5~1.5~ 2~ 1.5]^T$. It can be verified that $Bu_{\text{nom}}=v_{\text{nom}}$.
\\
\\
\textbf{Full information case:}
The saturation thresholds  $\Delta u^{min}$ and  $\Delta u^{max}$ are chosen so as to ensure $u(t)\in U$. This condition translates into $U_{\text{min}}\leq u_{\text{nom}}+ sat_{[\Delta u^{min},~\Delta u^{max}]}\leq U_{\text{max}}$. 
%
%
Denote $\mathbf{1}$ as a vector with all entries equal to 1.
For the instantaneous game a negative allocation/surplus is not allowed, so $U_{min}\geq 0\cdot\mathbf{1}$. Further, an allocation/surplus greater than the value of grand coalition is not allowed, so $U_{max}\leq v_{nom}(N) \cdot \mathbf{1}$. 
For the given game parameters, we see that 
the lower and upper thresholds for the saturation function are $-1$ and 
$5.5$, respectively. Next, we present the performance results of the robust control law given by equation \eqref{cl}. 
From Theorem \ref{thm1}, $\lim_{t\rightarrow \infty} {z}(t)$ converges to zero w.p.1 and as a result $\lim_{t\rightarrow \infty}
\frac{x(t)-x(0)}{t}$ converges to zero.
Fig. \ref{fig:subfig1} illustrates this behavior for the first component of coalition $\{1,2\}$.
Further, by Corollary \ref{cor:cor1}, the same control law ensures 
that the average allocations converge to the nominal allocations in the long run, in other words $\lim_{t\rightarrow \infty}\bar{a}(t)=a_{nom}$ and Fig. \ref{fig:subfig2} illustrates this behavior.
\begin{figure}[h]
\subfigure[Plot of $\frac{x_{\{1,2\}}(t)-x_{\{1,2\}}(0)}{t}$]{
\includegraphics[scale=0.525]{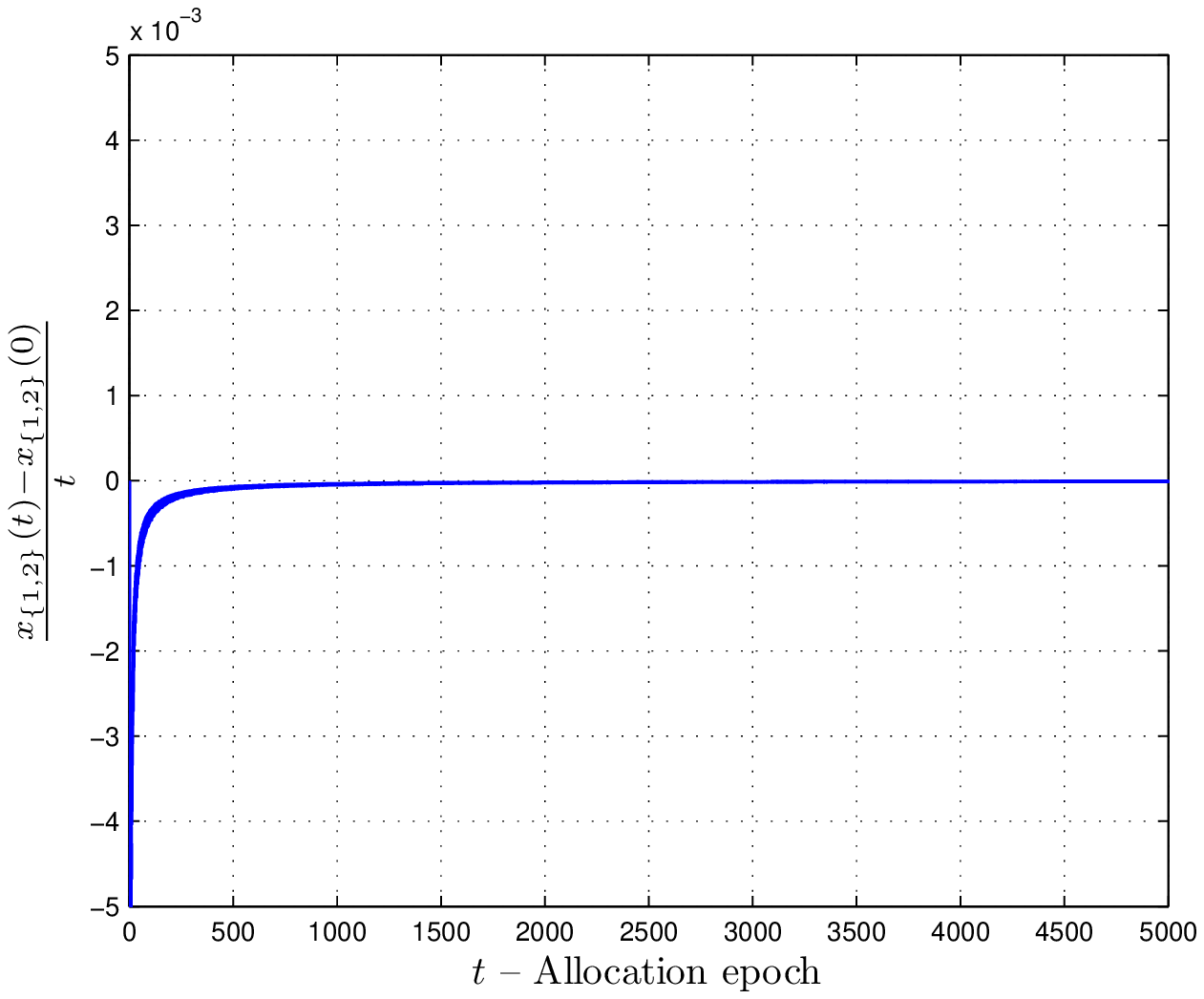}
\label{fig:subfig1}
}
\subfigure[Plot of $\lim_{t\rightarrow \infty}\bar{a}(t)-a_{nom}$]{
\includegraphics[scale=0.525]{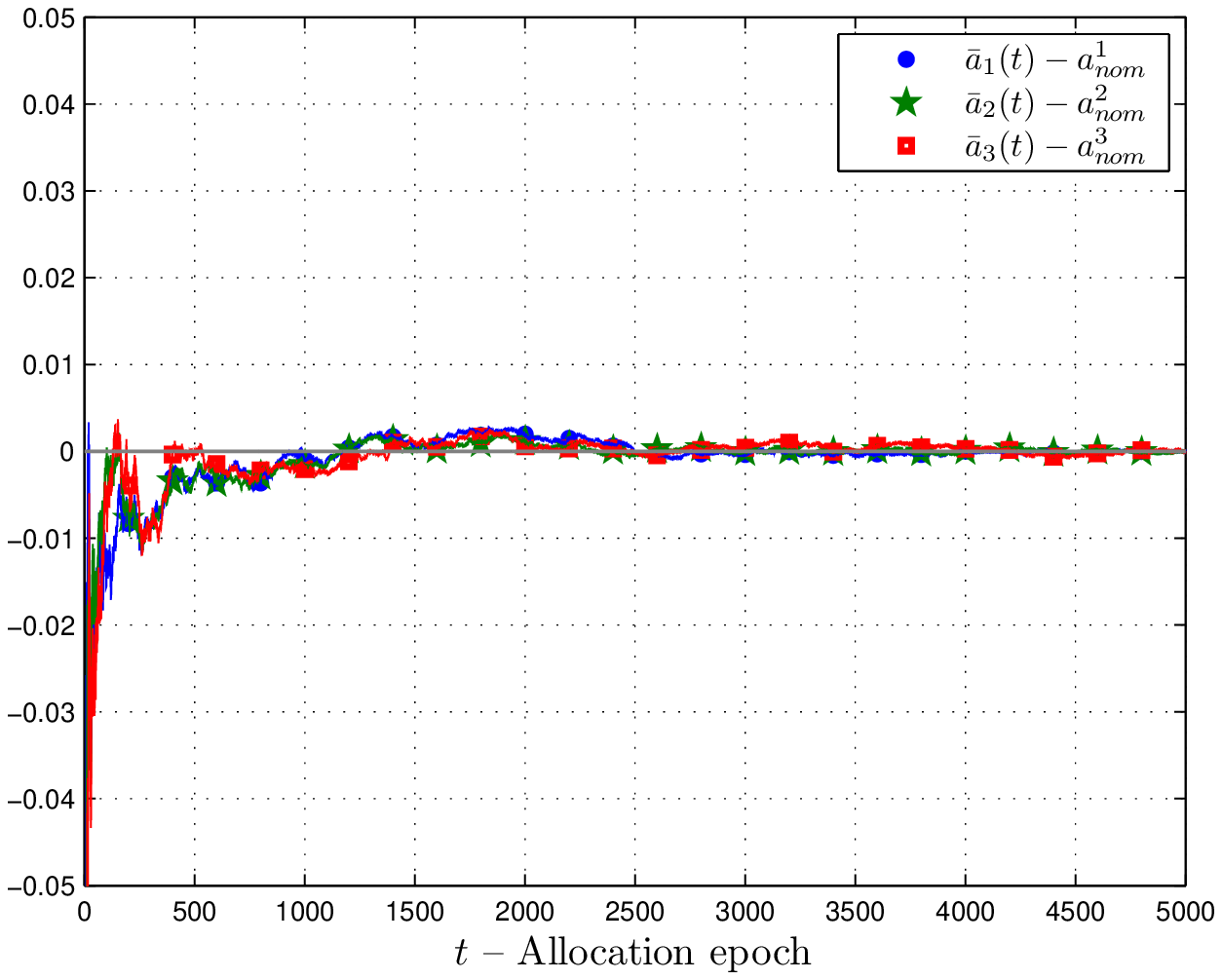}
\label{fig:subfig2}
}
\label{fig:subfigureExample}
\caption[Optional caption for list of figures]{Performance
of the control law given by \eqref{cl}.}
\end{figure}
%
%
\\
\noindent
\textbf{Partial information case:}
The choice of $\delta$ is crucial so as to ensure $u(t)\in U$. This condition translates to $U_{\text{min}}\leq u_{\text{nom}}+\delta B^\dagger sgn(x)\leq U_{\text{max}}$. We observe $-\sum_j |B^\dagger_{ij}|\leq \left(B^\dagger sgn(x)\right)_i\leq \sum_j |B^\dagger_{ij}|$. A conservative estimate of $\delta$ is obtained as
$U_{min}\leq {u_{nom}}\pm\delta\max_i\{\sum_j|B^\dagger_{ij}|\}\leq U_{max}$.
For $m=7$, we have $\max_i\{\sum_j|B^\dagger_{ij}|\}=2.11$. 
For the instantaneous game a negative allocation/surplus is not allowed, so $U_{min}\geq 0.\mathbf{1}$. Furthermore, an allocation/surplus greater than the value of grand coalition is not allowed, so $U_{max}\leq v_{nom}(N).\mathbf{1}$. 
We chose $\delta=1$, which satisfies the above stated requirements. 
\begin{figure}[h]
\subfigure[Plot of $\frac{x_{\{1,2\}}(t)-x_{\{1,2\}}(0)}{t}$]{
\includegraphics[scale=0.525]{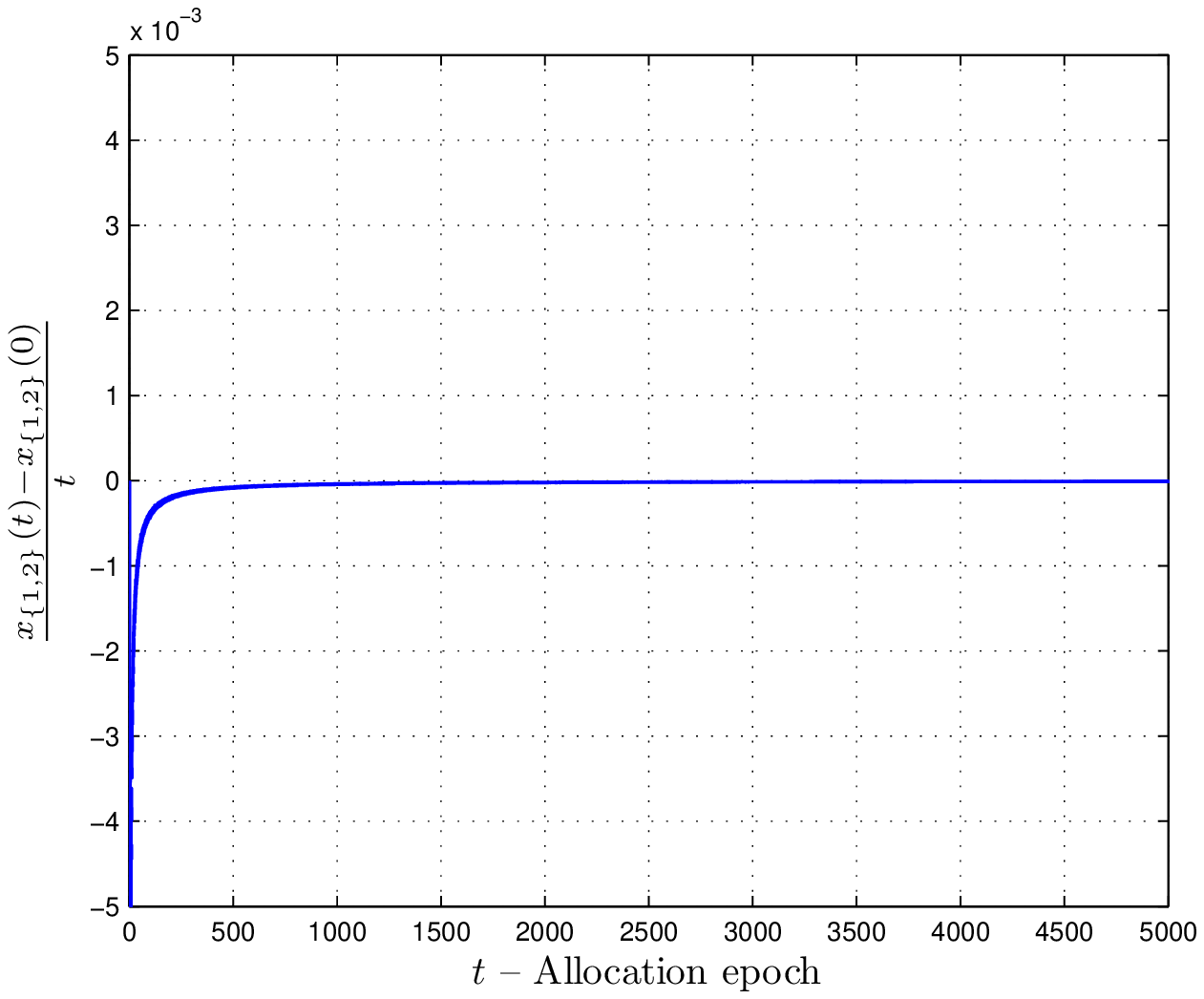}
\label{fig:subfig3}
}
\subfigure[Plot of $\lim_{t\rightarrow \infty}\bar{a}(t)-a_{nom}$ ]{
\includegraphics[scale=0.525]{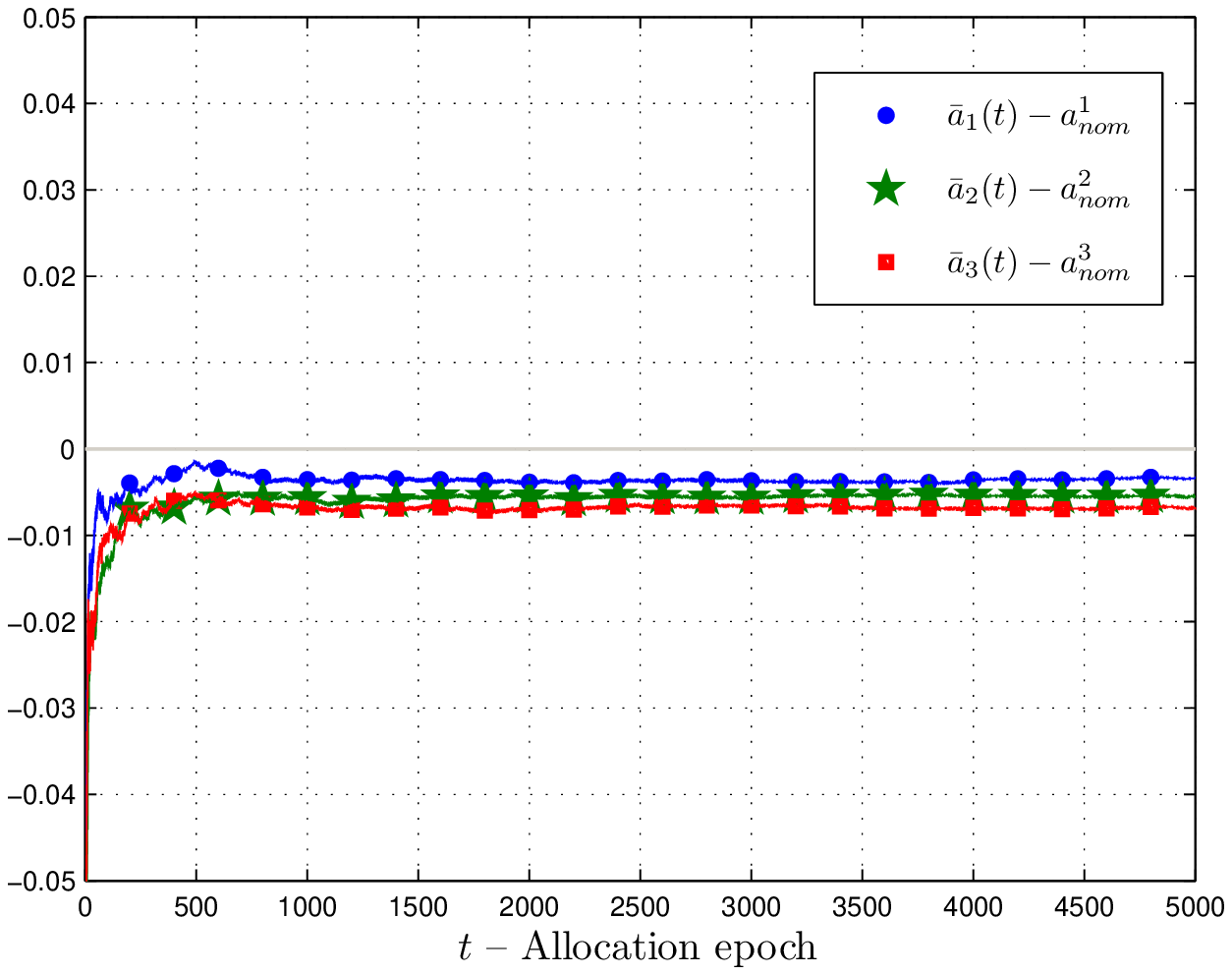}
\label{fig:subfig4}
}
\label{fig:subfigureExample}
\caption[Optional caption for list of figures]{Performance
of the control law given by \eqref{c2}.}
\end{figure}
Next, we present performance results of the robust control law given by equation (\ref{c2}). 
From Theorem \ref{thm2}, $x(t)$ converges to zero in probability 
with a specific choice of control law and as a result $\lim_{t\rightarrow \infty}
\frac{x(t)-x(0)}{t}$ converges to zero.
Fig. \ref{fig:subfig3} illustrates this behavior for the first component of coalition $\{1,2\}$. Further, by Corollary
\ref{cor:cor2}, the same control law ensures 
that the average allocations converge to the core $C(v_{nom})$ and from equation \eqref{c2} it is clear that the instantaneous
allocations lie in a neighborhood of nominal allocations.
As a result there is uncertainty in the convergence of
average allocations towards nominal allocations on the long run and
Fig. \ref{fig:subfig4} illustrates this behavior. 
%
%
%
\section{Conclusions}
\label{sec:conclusions}
In this paper we studied dynamic cooperative games where at each instant of time 
the value of each coalition of players is unknown but varies within a bounded polyhedron.
With the assumption that the average value of each coalition in the long run is known with
certainty, we presented robust allocations schemes, which converge to the core, under two informational settings. 
 We proved the convergence of both allocation rules using
Lyapunov stochastic stability theory. Furthermore, we established connections of Lyapunov stability theory to concepts of approachability and attainability. The control laws
or allocation schemes are derived on the premise that the GD knows a priori, the nominal
allocation vector. If this information is not available then the problem can be treated as
a learning process where the GD is trying to learn the (balanced) nominal game from the
instantaneous games. The allocation rules designed in this paper assure stability of the
coalitions in average, and as a result capture patience and expectations of the players in an
integral sense. The modeling aspects of generic dynamic coalitional games are open questions at this point of time.
%
%
\bibliographystyle{elsarticle-num}

%
%
%
%
\end{document}